\documentclass[apjl]{emulateapj}
\usepackage{psfig,amsfonts,amsmath,graphicx,natbib,apjfonts,lscape}

\def\ra#1#2#3{#1$^{\rm h}$#2$^{\rm m}$#3$^{\rm s}$}
\def\dec#1#2#3{$#1^\circ#2'#3''$}
\def\nod{\nodata}

\shorttitle{}
\shortauthors{Leibler \& Berger}

\def\cfa{1}

\begin{document}

\title{The Stellar Ages and Masses of Short GRB Host Galaxies:
Investigating the Progenitor Delay Time Distribution and the Role of
Mass and Star Formation in the Short GRB Rate}

\author{ 
C.~N.~Leibler\altaffilmark{\cfa} and
E.~Berger\altaffilmark{\cfa}
}

\altaffiltext{\cfa}{Harvard-Smithsonian Center for Astrophysics, 60
Garden Street, Cambridge, MA 02138}

\begin{abstract} We present multi-band optical and near-infrared
observations of 19 short $\gamma$-ray burst (GRB) host galaxies, aimed
at measuring their stellar masses and population ages.  The goals of
this study are to evaluate whether short GRBs track the stellar mass
distribution of galaxies, to investigate the progenitor delay time
distribution, and to explore any connection between long and short GRB
progenitors.  Using single stellar population models we infer masses
of ${\rm log} (M_*/M_\odot)\approx 8.8-11.6$, with a median of
$\langle{\rm log} (M_*/M_\odot)\rangle\approx 10.1$, and population
ages of $\tau_* \approx 0.03-4.4$ Gyr with a median of
$\langle\tau_*\rangle\approx 0.3$ Gyr.  We further infer maximal
masses of ${\rm log}(M_*/M_\odot)\approx 9.7-11.9$ by assuming stellar
population ages equal to the age of the universe at each host's
redshift.  Comparing the distribution of stellar masses to the general
galaxy mass function we find that short GRBs track the cosmic stellar
mass distribution only if the late-type hosts generally have maximal
masses.  However, there is an apparent dearth of early-type hosts
compared to the equal contribution of early- and late-type galaxies to
the cosmic stellar mass budget.  These results suggest that stellar
mass may not be the sole parameter controlling the short GRB rate, and
raise the possibility of a two-component model with both mass and star
formation playing a role (reminiscent of the case for Type Ia
supernovae).  If short GRBs in late-type galaxies indeed track the
star formation activity, the resulting typical delay time is $\sim
0.2$ Gyr, while those in early-type hosts have a typical delay of
$\sim 3$ Gyr.  Using the same stellar population models we fit the
broad-band photometry for 22 long GRB host galaxies in a similar
redshift range and find that they have significantly lower masses and
younger population ages, with $\langle{\rm log}(M_*/ M_\odot)\rangle
\approx 9.1$ and $\langle\tau_*\rangle\approx 0.06$ Gyr, respectively;
their maximal masses are similarly lower, $\langle{\rm log}
(M_*/M_\odot)\rangle\approx 9.6$, and as expected do not track the
galaxy mass function.  Most importantly, the two GRB host populations
remain distinct even if we consider only the star-forming hosts of
short GRBs, supporting our previous findings (based on star formation
rates and metallicities) that the progenitors of long GRBs and short
GRBs in late-type galaxies are distinct.  Given the much younger
stellar populations of long GRB hosts (and hence of long GRB
progenitors), and the substantial differences in host properties, we
caution against the use of Type I and II designations for GRBs since
this may erroneously imply that all GRBs which track star formation
activity share the same massive star progenitors.  \end{abstract}


\section{Introduction}
\label{sec:intro}

The progenitors of the short-duration $\gamma$-ray bursts (GRBs)
remain unidentified at the present.  However, in recent years
important constraints have been placed on their nature based on
observations of the prompt emission, afterglows, and host galaxies.
In particular, the discovery of short GRBs in elliptical galaxies
\citep{bpc+05,bpp+06} demonstrated that at least some of the
progenitors belong to an old stellar population ($\gtrsim 1$ Gyr).  On
the other hand, the bulk of the identified host population exhibits
star formation activity \citep{bfp+07}, albeit with lower star
formation rates and higher metallicities than the hosts of long GRBs
\citep{ber09}.  On a sub-galactic scale, the locations of short GRBs
do not coincide with regions of active star formation
\citep{ffp+05,fbf10,rwl+10}, and instead trace the rest-frame optical
light distribution, indicative of a connection with older stellar
populations compared to long GRBs \citep{fbf10,ber10b}.  Similarly,
the offsets of short GRBs are substantially larger than those of long
GRBs \citep{fbf10}, with recent potential evidence for a population of
short bursts with offsets of tens of kpc, possibly due to natal kicks
\citep{ber10}.

The afterglow and prompt emission properties likewise point to a
distinct origin for the two GRB types.  Short GRBs have significantly
fainter optical and X-ray afterglows than those of long GRBs
\citep{nak07,gbb+08,kkz+08,bcf+09,nfp+09,ber10}, although they exhibit
similar flux ratios between the two bands \citep{nfp+09,ber10}.  Their
isotropic-equivalent $\gamma$-ray energies, $E_{\rm\gamma,iso}\approx
10^{49}-10^{52}$ erg, are also generally lower than for long GRBs
\citep{bfp+07,gbb+08,nfp+09,ber10}.  These results point to an overall
lower energy budget, and potentially lower circumburst densities
\citep{sbk+06,ber10}.  Taken in conjunction, the emission and
environmental properties point to a distinct origin for short and long
GRBs, and indicate that the short GRB progenitors are not dominated by
a young population (e.g., massive stars or young magnetars).

Despite this crucial insight three basic and inter-related questions
remain open at the present, whose answers can shed light on the nature
of the progenitors:
\begin{itemize}
\item Does the short GRB rate depend solely on the stellar masses of
the host galaxies, or does star formation play a role?
\item What is the delay time distribution of short GRBs?
\item Is there an overlap between long and short GRBs, such that
duration is not uniquely mapped to progenitor?
\end{itemize}
In the context of the first two questions, the popular compact object
binary progenitors (NS-NS or NS-BH; \citealt{elp+89,pac91,npp92}) are
expected to track stellar mass if the merger timescales are long
($\gtrsim {\rm Gyr}$).  On the other hand, if a substantial rapid
merger channel exists (e.g., \citealt{bpb+06}), then star formation
activity will also impact the short GRB rate.  Similarly, the
distribution of merger timescales will impact the mix of early- and
late-type hosts \citep{zr07}.  In particular, if mass is the dominant
factor we expect a roughly equal number of short GRBs in early- and
late-type galaxies\footnotemark\footnotetext{This assumes that there
are no systematic differences between short GRBs in early- and
late-type galaxies that would influence their detectability, such as
the typical ambient density or energy release.} since at the relevant
redshift range each galaxy population type accounts for about half of
the cosmic stellar mass budget (e.g., \citealt{isl+10}); a connection
with star formation will serve to increase the late-type fraction.
The same trends will be apparent for any progenitor system with a
broad range of delay times, such as accretion-induced collapse (AIC)
of neutron stars \citep{qwc+98}, and delayed magnetar formation
through binary white dwarf mergers or white dwarf AIC
\citep{lwc+06,mqt08}.

Related to the questions of delay times and a connection to stellar
mass or star formation is whether durations and progenitors are
directly linked, i.e., whether some short GRBs arise from massive star
progenitors like long GRBs (e.g., \citealt{vzo+09,lmb10}).  If such a
subset of events exists, it will be manifested in a population of
short GRBs with negligible delay times of $\sim 10$ Myr, or
equivalently hosts with very young stellar populations and vigorous
star formation activity.  Such stellar populations are clearly present
in long GRB hosts \citep{chg04,lbk+10}.  In this context,
\citet{zzl+07} and \citet{zzv+09} proposed a re-classification of long
and short GRBs into Type I and II events associated with old and young
stellar populations, respectively.  Building on this concept,
\citet{vzo+09} recently carried out a joint analysis of the short GRB
luminosity-redshift distribution and the BATSE ${\rm log}N-{\rm log}S$
distribution and concluded that $\sim 90\%$ of the bursts classified
as short have negligible delay times and therefore directly track the
cosmic star formation history.  These authors therefore conclude that
most short GRBs have massive star progenitors similar to those of long
GRBs.

Here we take a different approach to exploring the relation between
short GRBs and stellar mass, the short GRB delay time distribution,
the possibility that short GRBs also track star formation, and whether
such a connection indicates a relation to long GRB progenitors.  Our
study relies on multi-band optical and near-infrared observations of
short GRB host galaxies from which we extract the relevant stellar
population properties: mass and age.  We confront the resulting mass
distribution with the general galaxy stellar mass function to assess
the relation to mass for the population as a whole, and separately for
the early- and late-type hosts.  We also use the inferred stellar
population ages to place constraints on the typical delay times in
early- and late-type hosts.  Furthermore, we uniformly re-analyze the
optical/near-IR spectral energy distributions of long GRB hosts to
allow a direct comparison with the inferred properties of short GRB
hosts.  This allows for a comparison of the stellar populations in the
two host samples with greater diagnostic power than the use of
secondary indicators such as the luminosity-redshift or ${\rm
log}N-{\rm log}S$ distributions.

The plan of the paper is as follows.  The sample properties and
optical/near-IR observations are summarized in \S\ref{sec:obs}.  The
stellar population models and fitting procedure are described in
\S\ref{sec:model}.  In \S\ref{sec:res} we describe the resulting
distributions of stellar masses and ages for a range of models and
assumptions.  We confront the mass distribution of short GRB hosts
with the galaxy mass function in \S\ref{sec:mass}, and compare the
stellar population properties of long and short GRB hosts in
\S\ref{sec:long}.  We draw conclusions about the delay times of short
GRB progenitors in the context their relation to mass and star
formation in \S\ref{sec:delay}, and summarize our key findings in
\S\ref{sec:conc}.

\section{Short GRB Sample and Observations}
\label{sec:obs}

The sample of short GRB hosts described in this paper includes 19
galaxies, of which 9 have been identified in coincidence with optical
afterglow positions
\citep{bpc+05,ffp+05,hwf+05,cmi+06,sbk+06,bcf+09,dmc+09,gfl+09,mkr+10,rot+10},
8 have been identified in coincidence with {\it Swift}/X-ray Telescope
(XRT) error circles (e.g., \citealt{bpp+06,bfp+07}), and 2 (GRBs
070809 and 090515) have been identified as putative hosts based on
chance coincidence probabilities for bursts with optical afterglows
and no coincident host galaxies \citep{ber10}.  Two of the hosts
(051210 and 070729) do not have spectroscopic redshift measurements,
and we infer photometric redshifts as part of the stellar population
analysis presented here (\S\ref{sec:model}).  In the analysis below we
address the properties and implications of the full sample, as well as
the subset of secure host associations (i.e., those found in
coincidence with optical afterglow positions).  The properties of the
bursts are summarized in Table~\ref{tab:data}.  We note that a few
additional secure hosts have photometric measurements in only $1-2$
optical filters and no spectroscopic redshifts (e.g., GRBs 060121 and
060313); these do not allow us to infer the required photometric
redshifts, stellar masses, and ages with any confidence.

\subsection{Optical and Near-IR Observations}

We obtained multi-band optical and near-IR observations that cover
rest wavelengths across the Balmer break and in the near-IR.  This
allows us to robustly assess the stellar population ages and the
stellar masses.  Optical observations were obtained with the Low
Dispersion Survey Spectrograph (LDSS3) on the Magellan/Clay 6.5-m
telescope, and the Gemini Multi-Object Spectrograph (GMOS;
\citealt{hja+04}) on the Gemini North and South 8-m telescopes.
Near-IR observations were performed with the Persson's Auxiliary
Nasmyth Infrared Camera (PANIC) on the Magellan/Baade 6.5-m telescope.

\begin{figure}
\includegraphics[angle=0,width=1.725in]{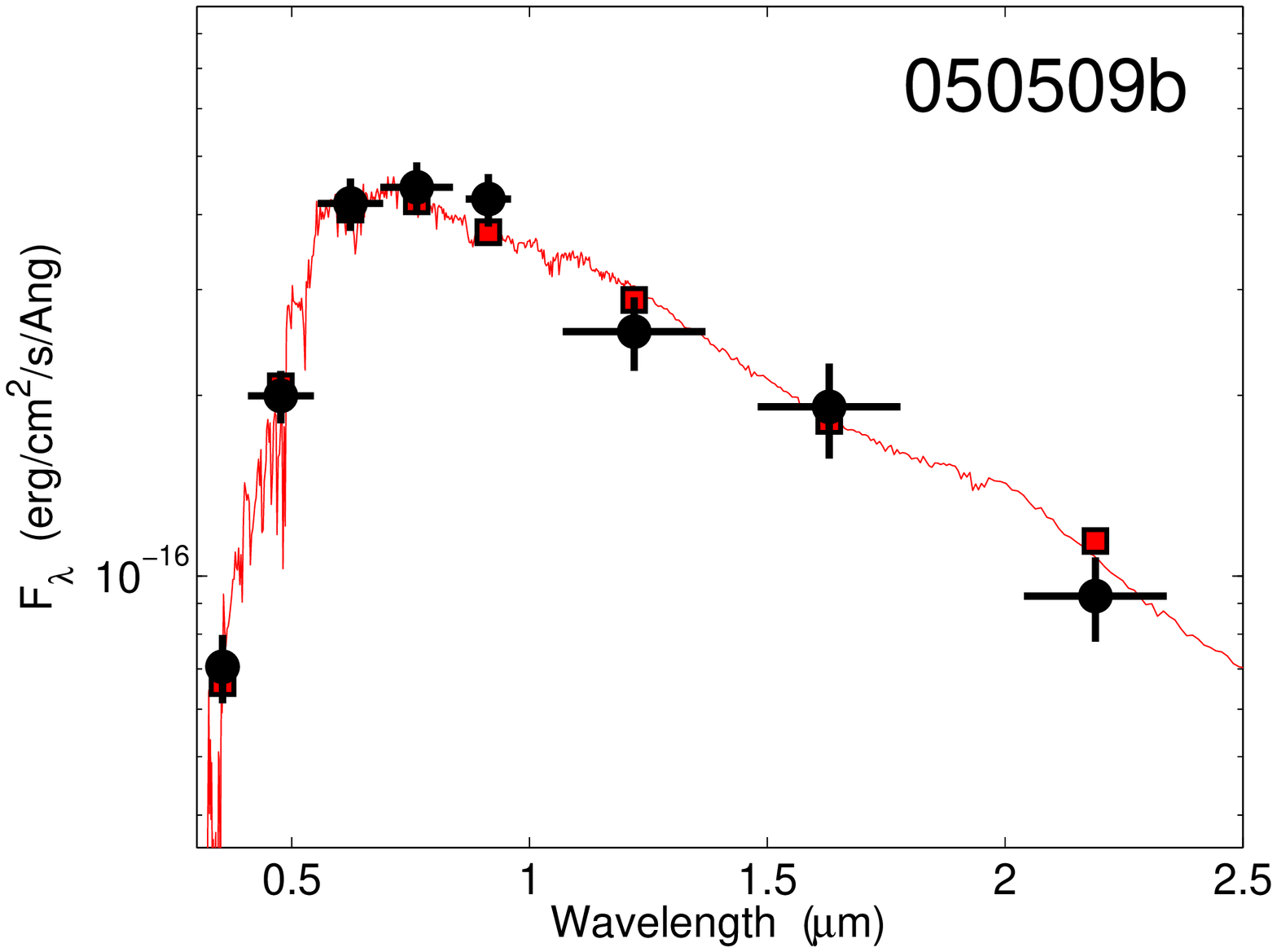}
\includegraphics[angle=0,width=1.725in]{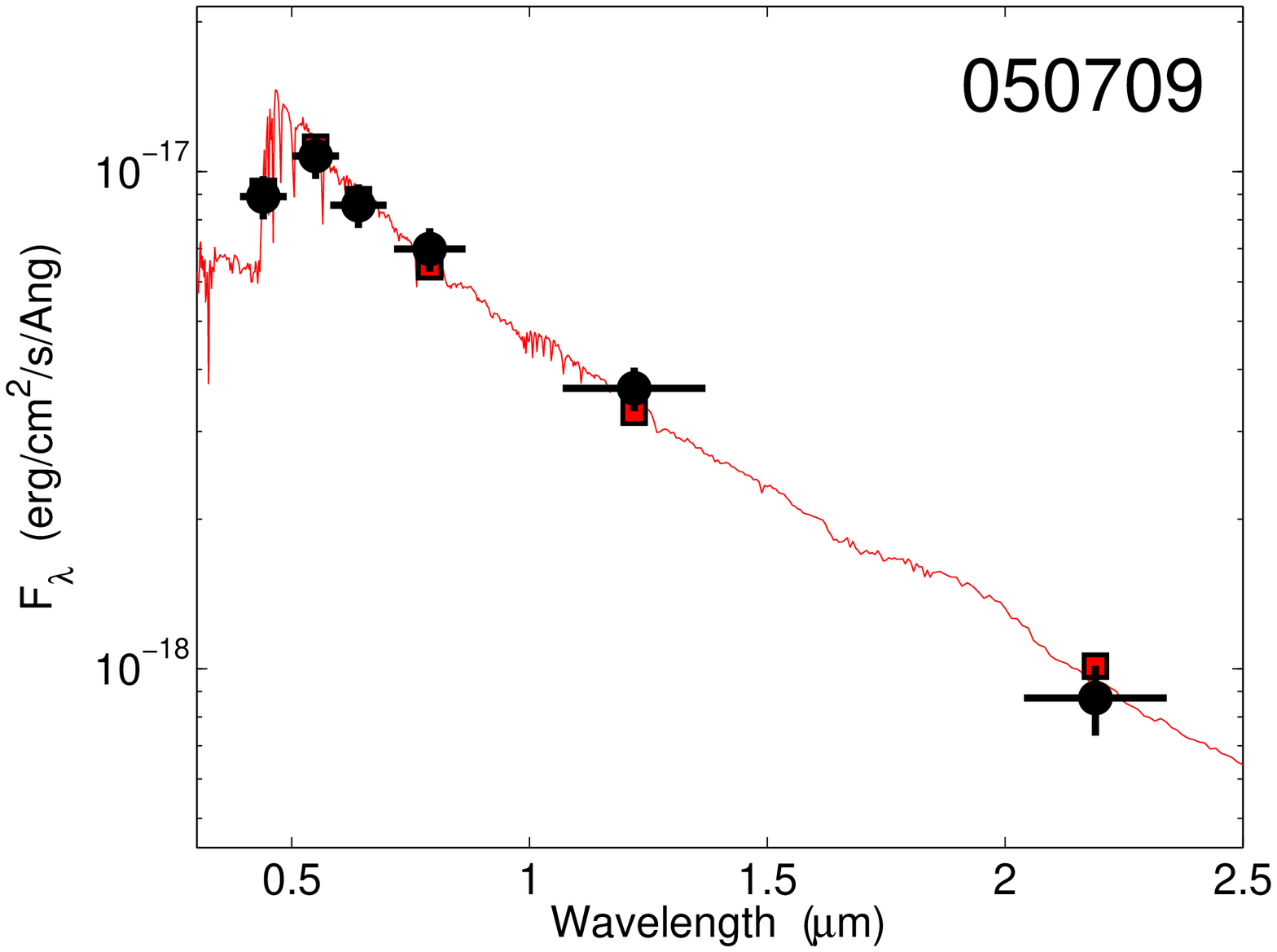}
\includegraphics[angle=0,width=1.725in]{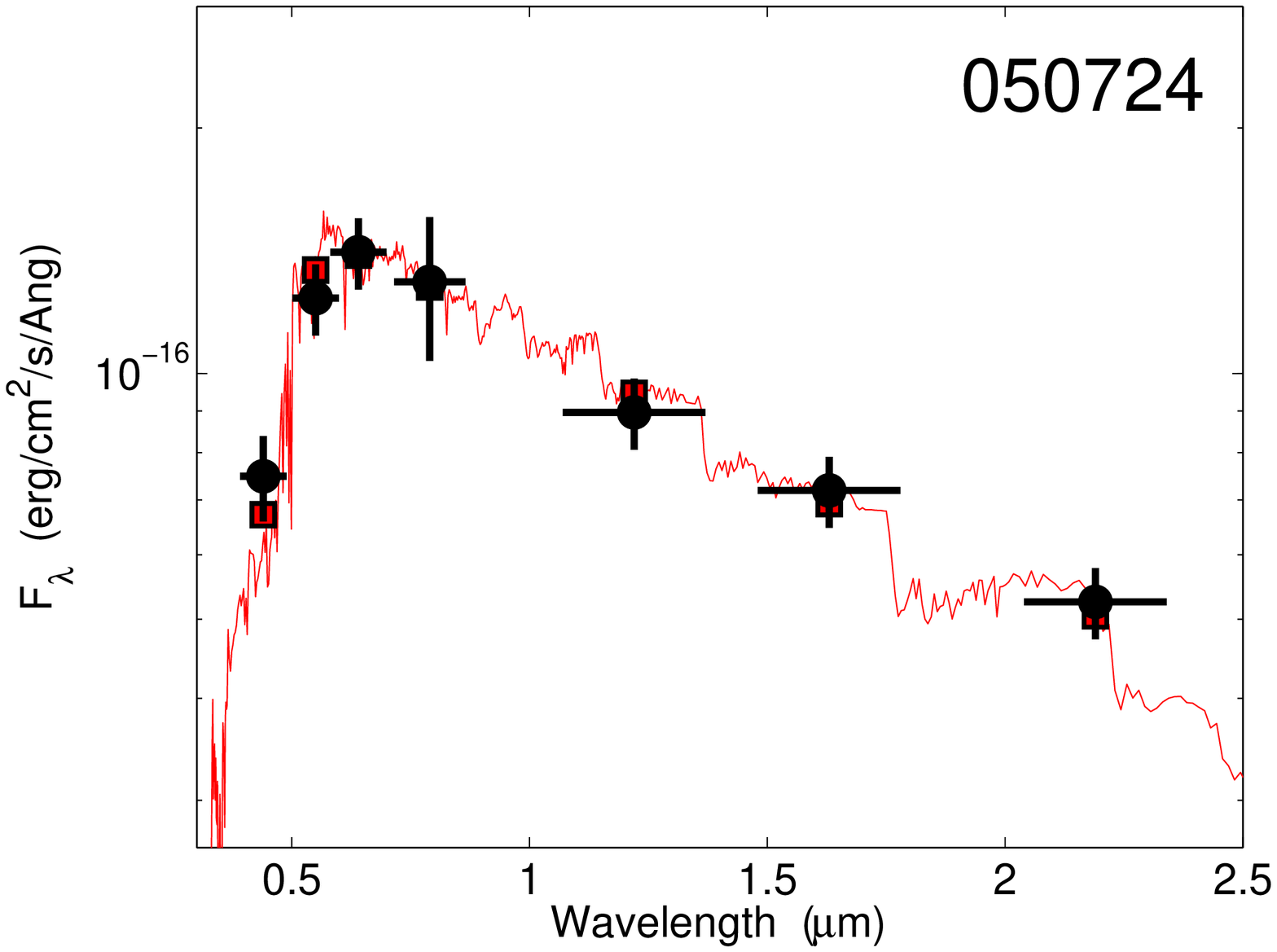}
\includegraphics[angle=0,width=1.725in]{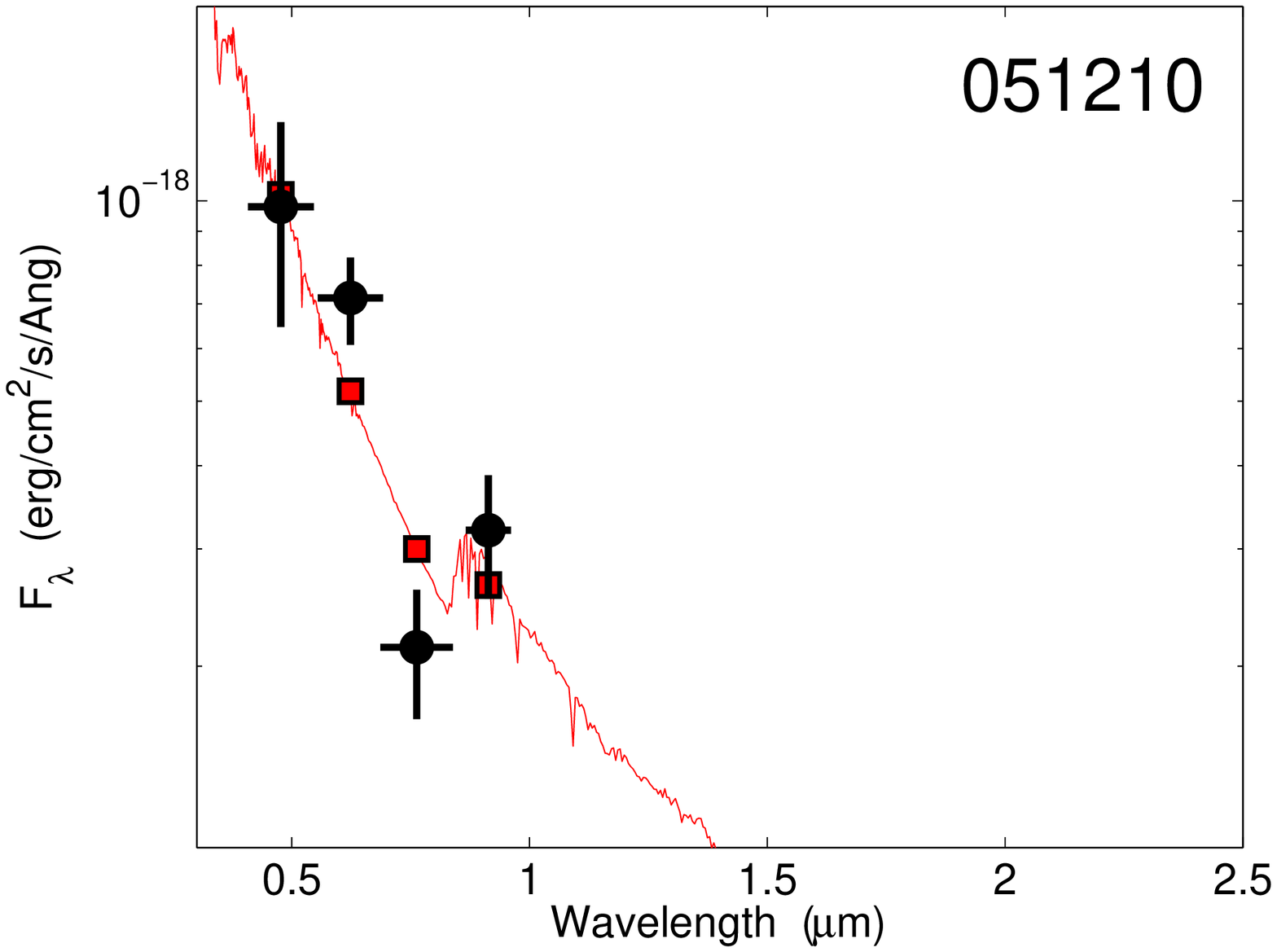}\\
\includegraphics[angle=0,width=1.725in]{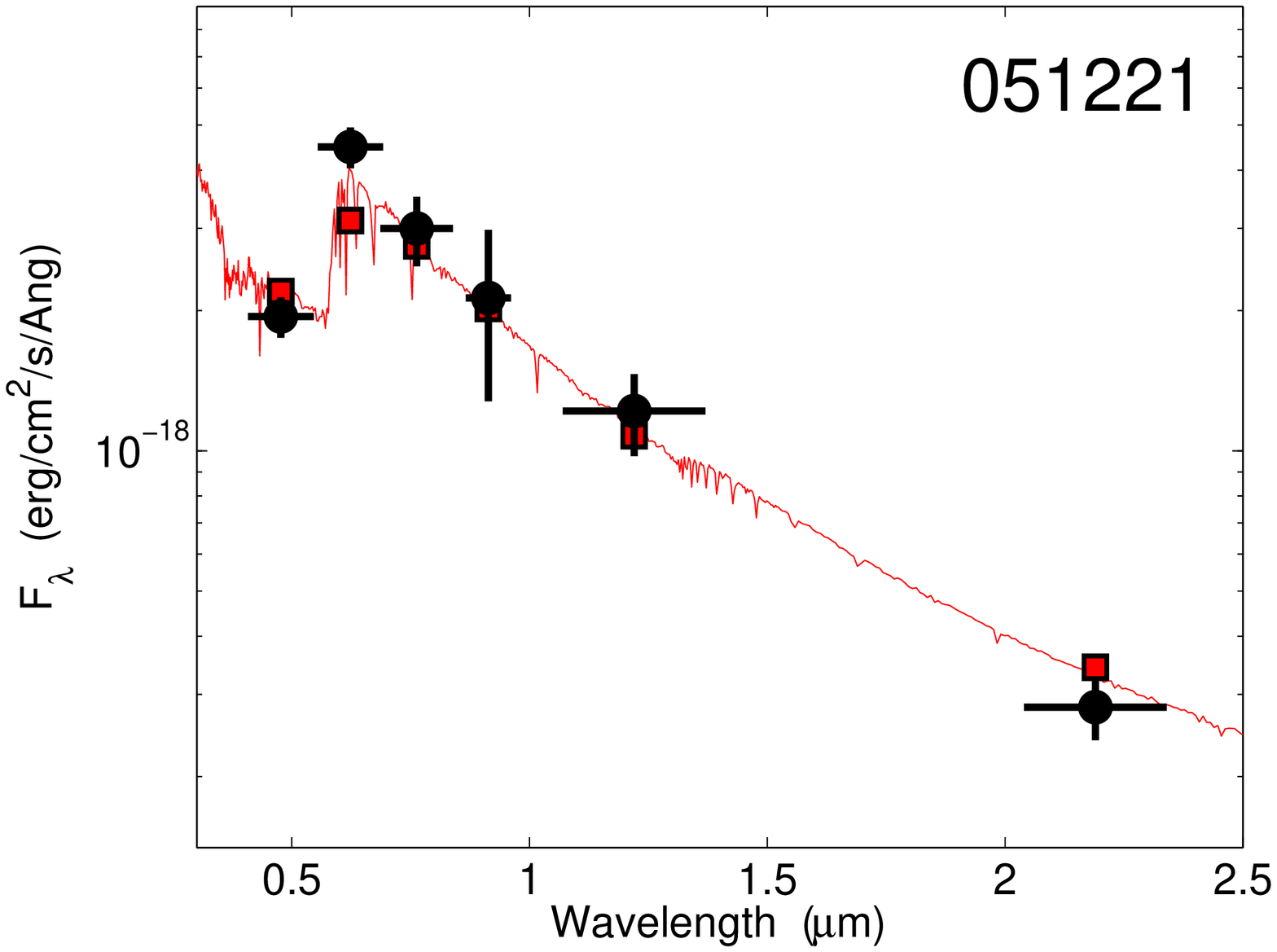}
\includegraphics[angle=0,width=1.725in]{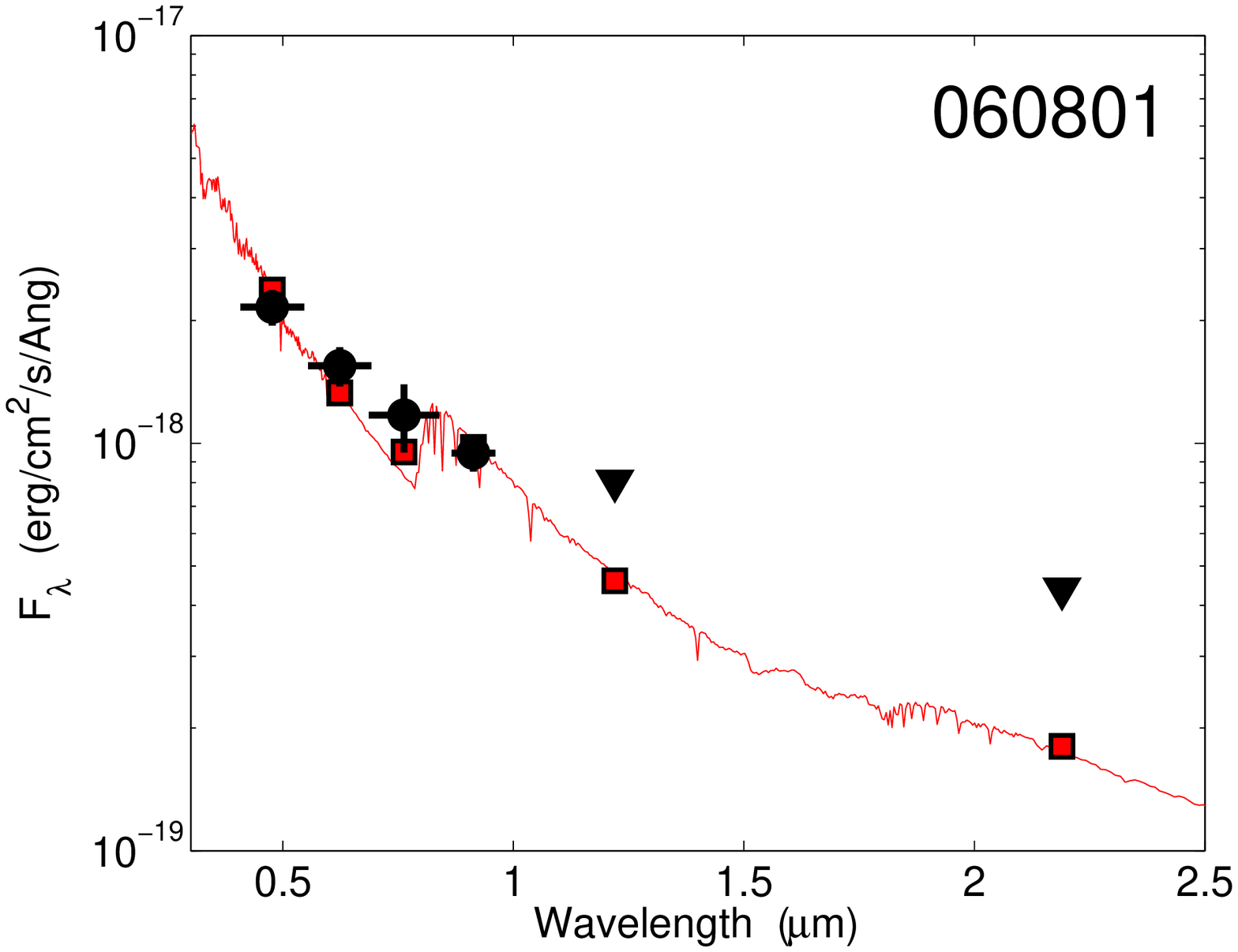}
\includegraphics[angle=0,width=1.725in]{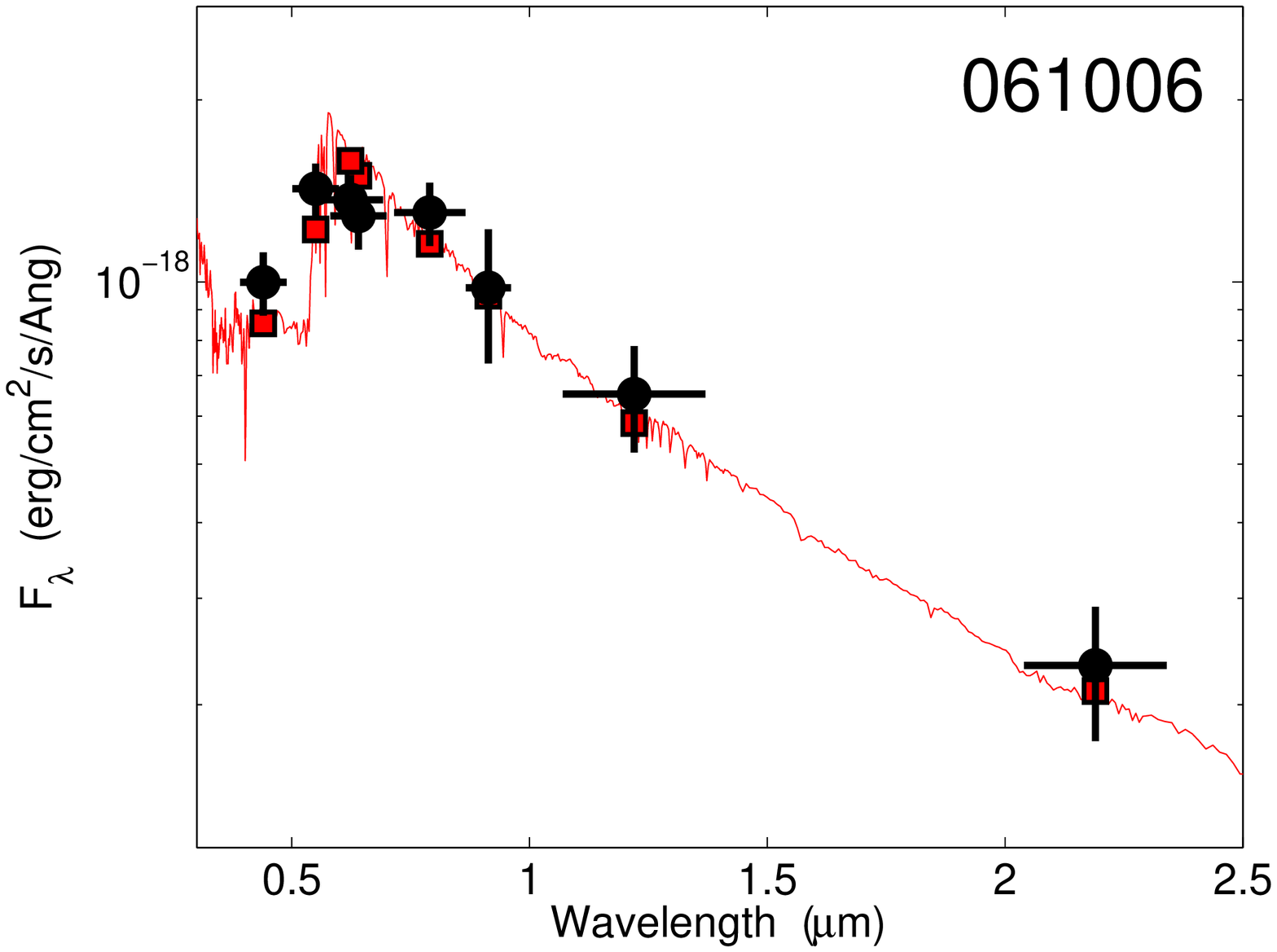}
\includegraphics[angle=0,width=1.725in]{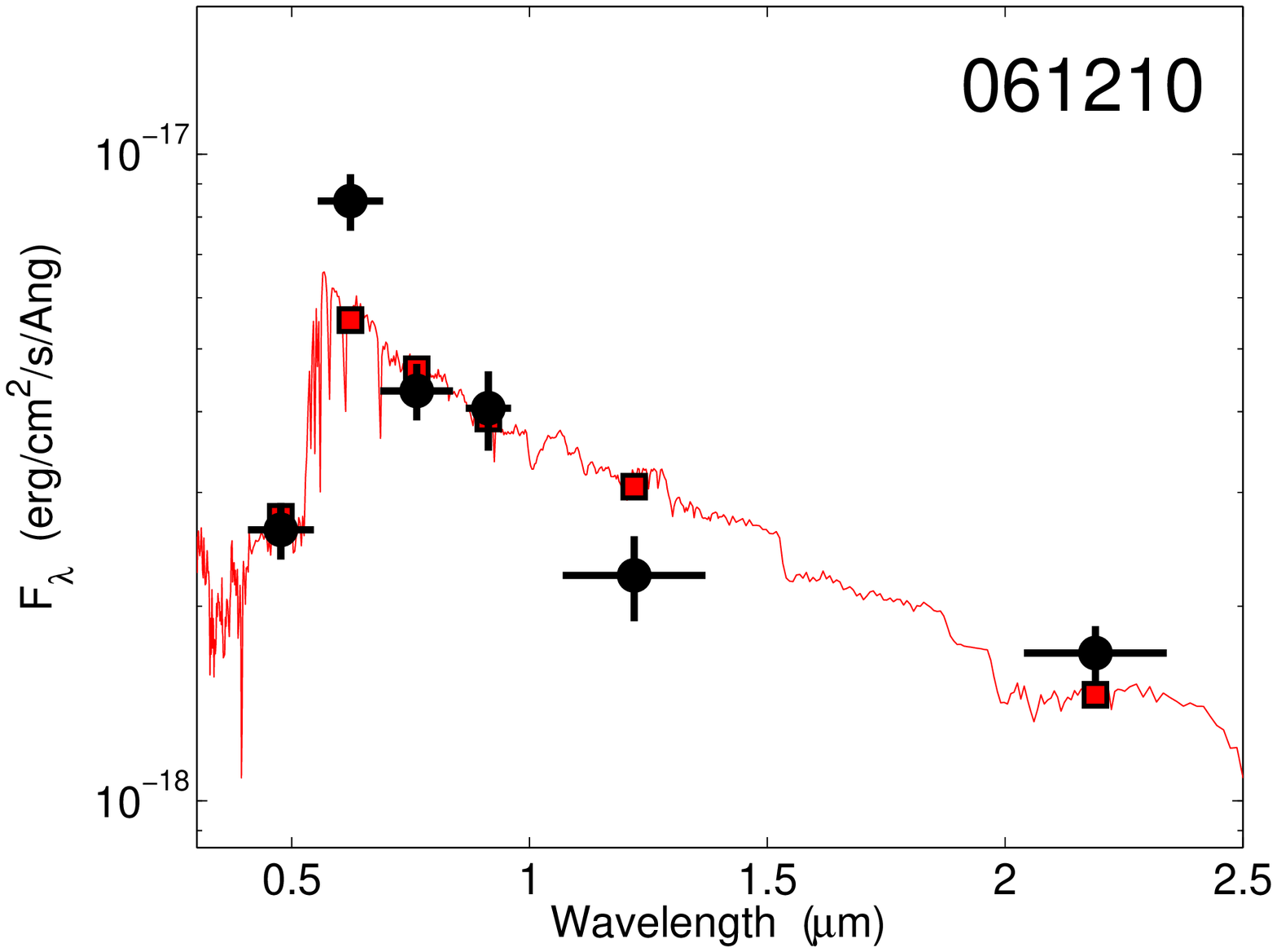}\\
\includegraphics[angle=0,width=1.725in]{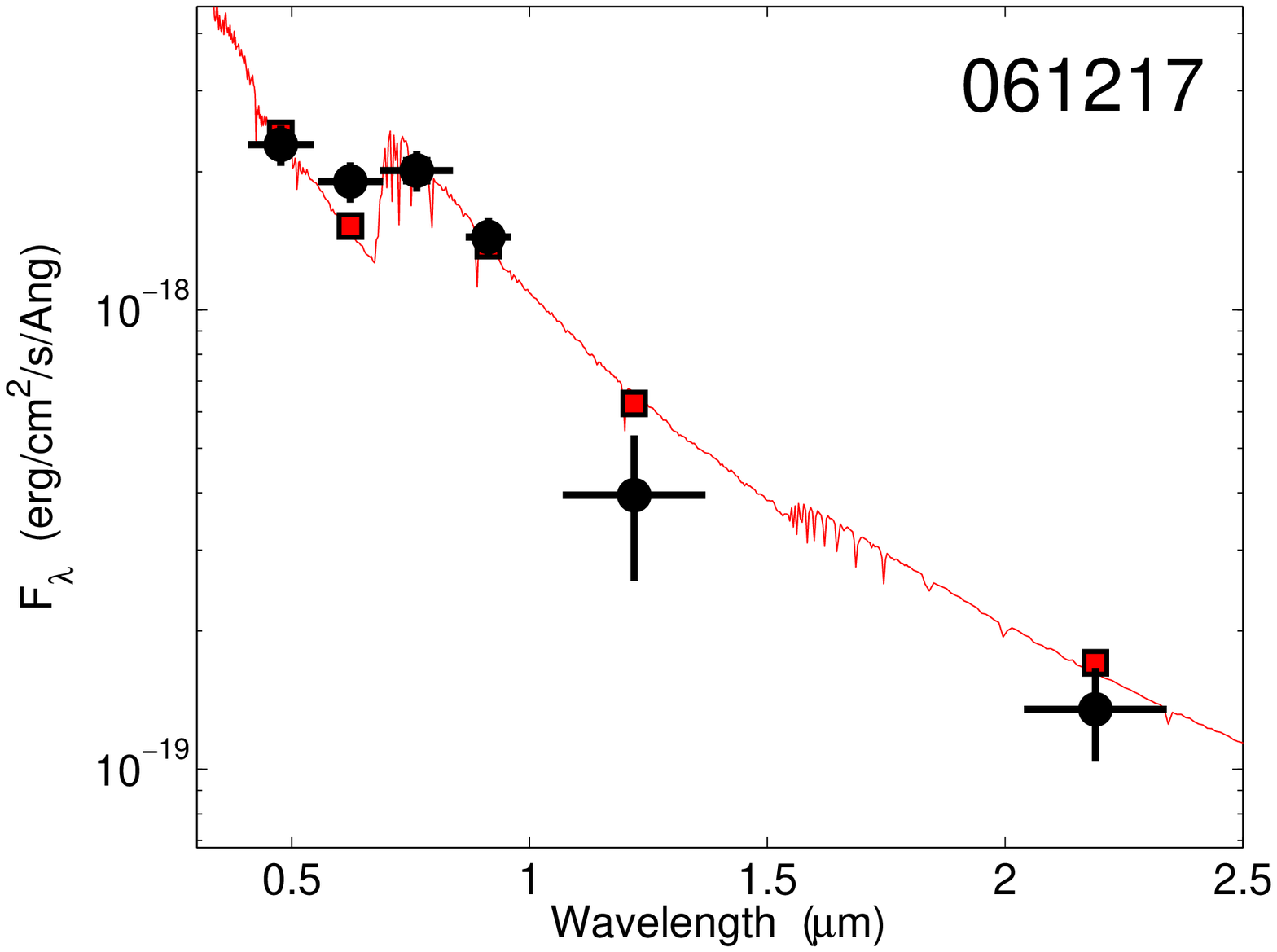}
\includegraphics[angle=0,width=1.725in]{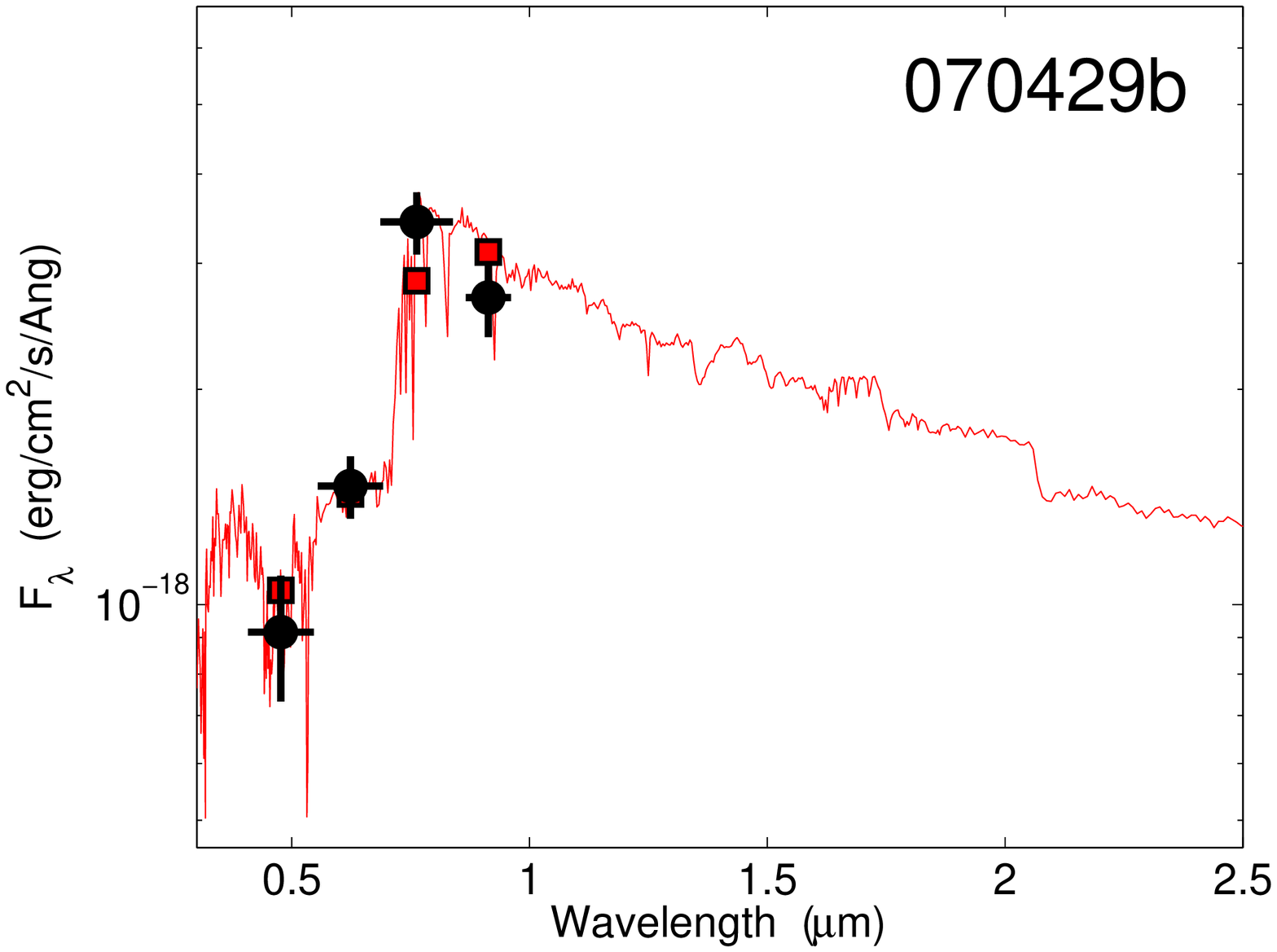}
\includegraphics[angle=0,width=1.725in]{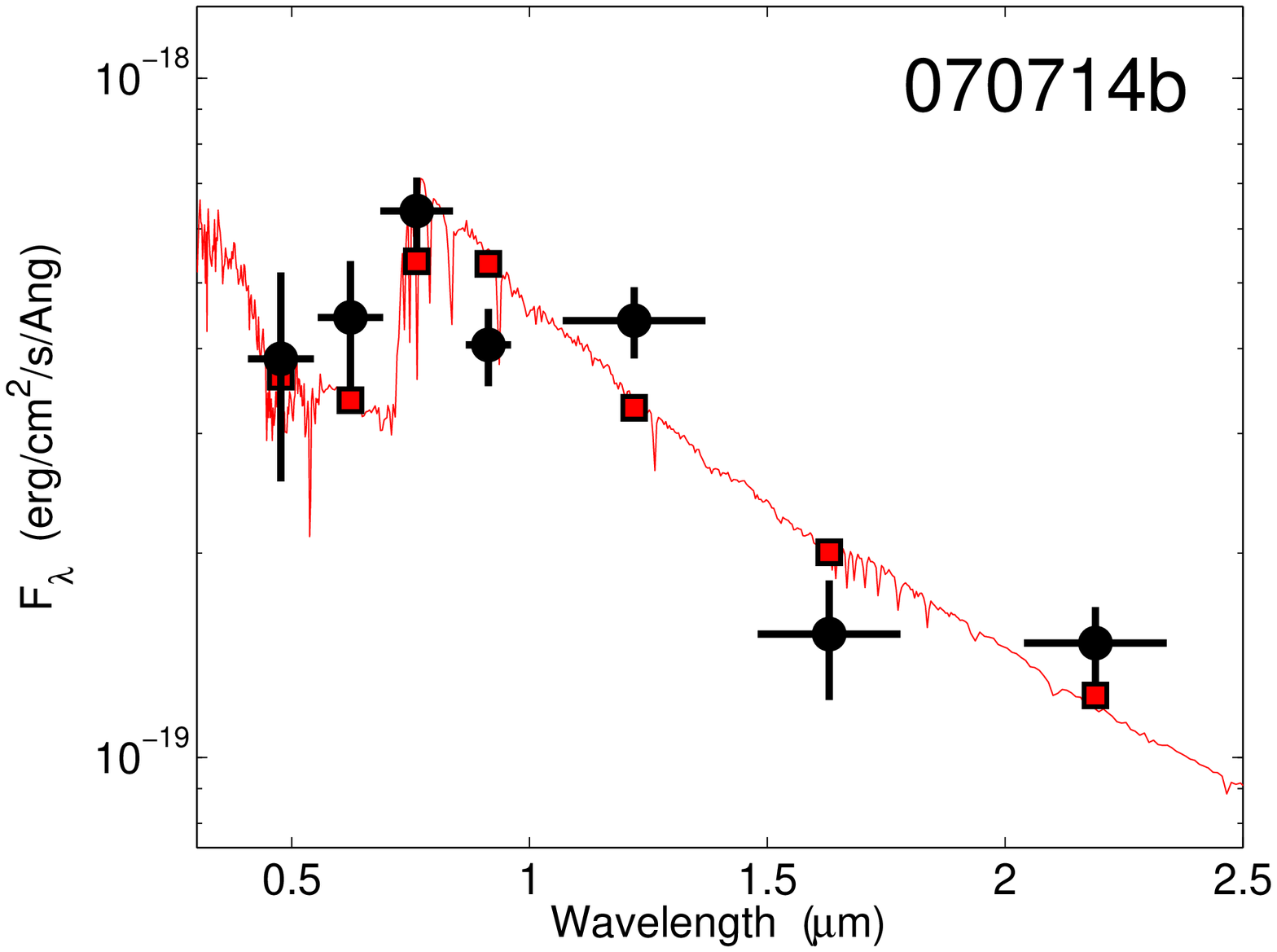}
\includegraphics[angle=0,width=1.725in]{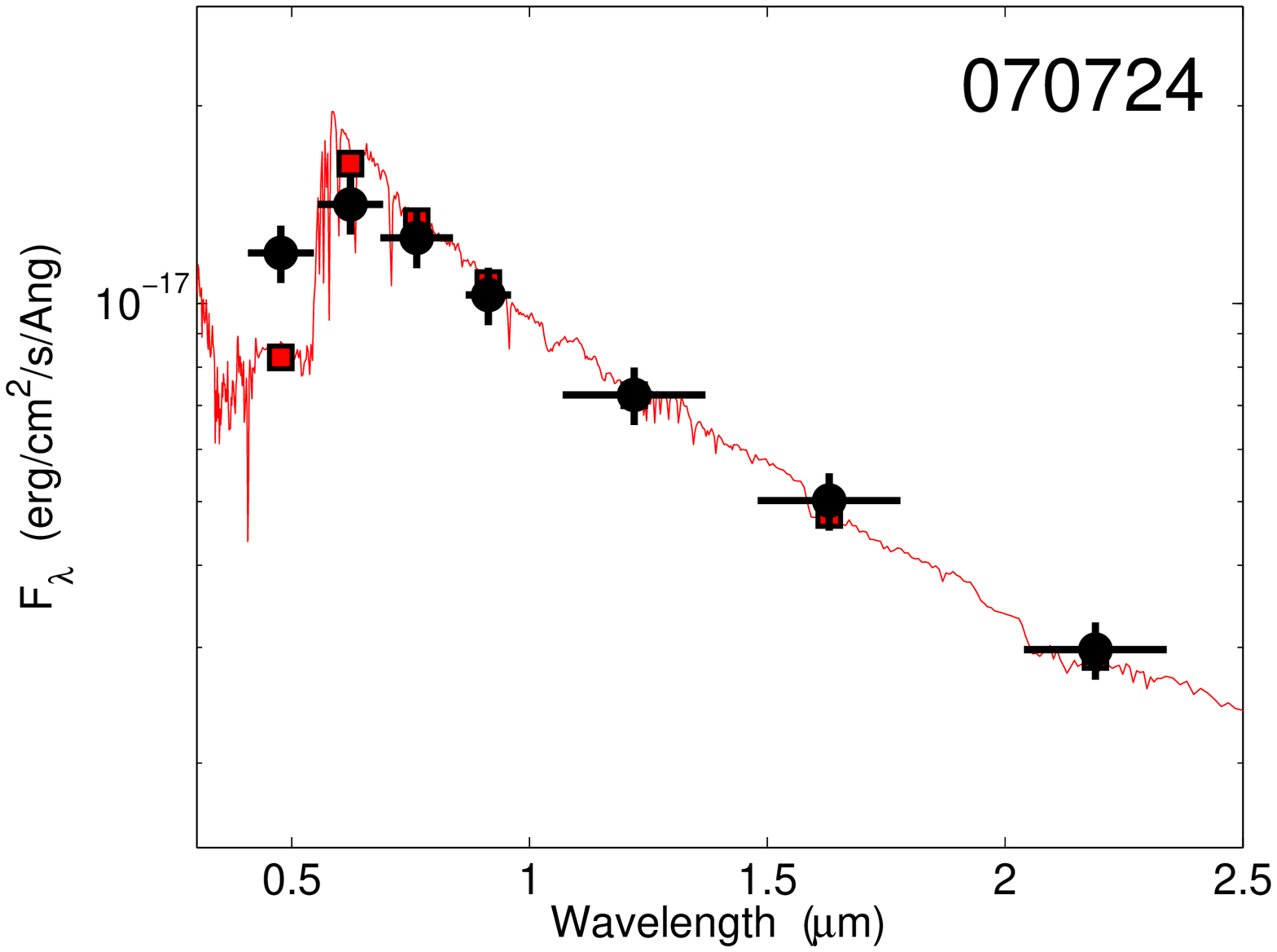}\\
\includegraphics[angle=0,width=1.725in]{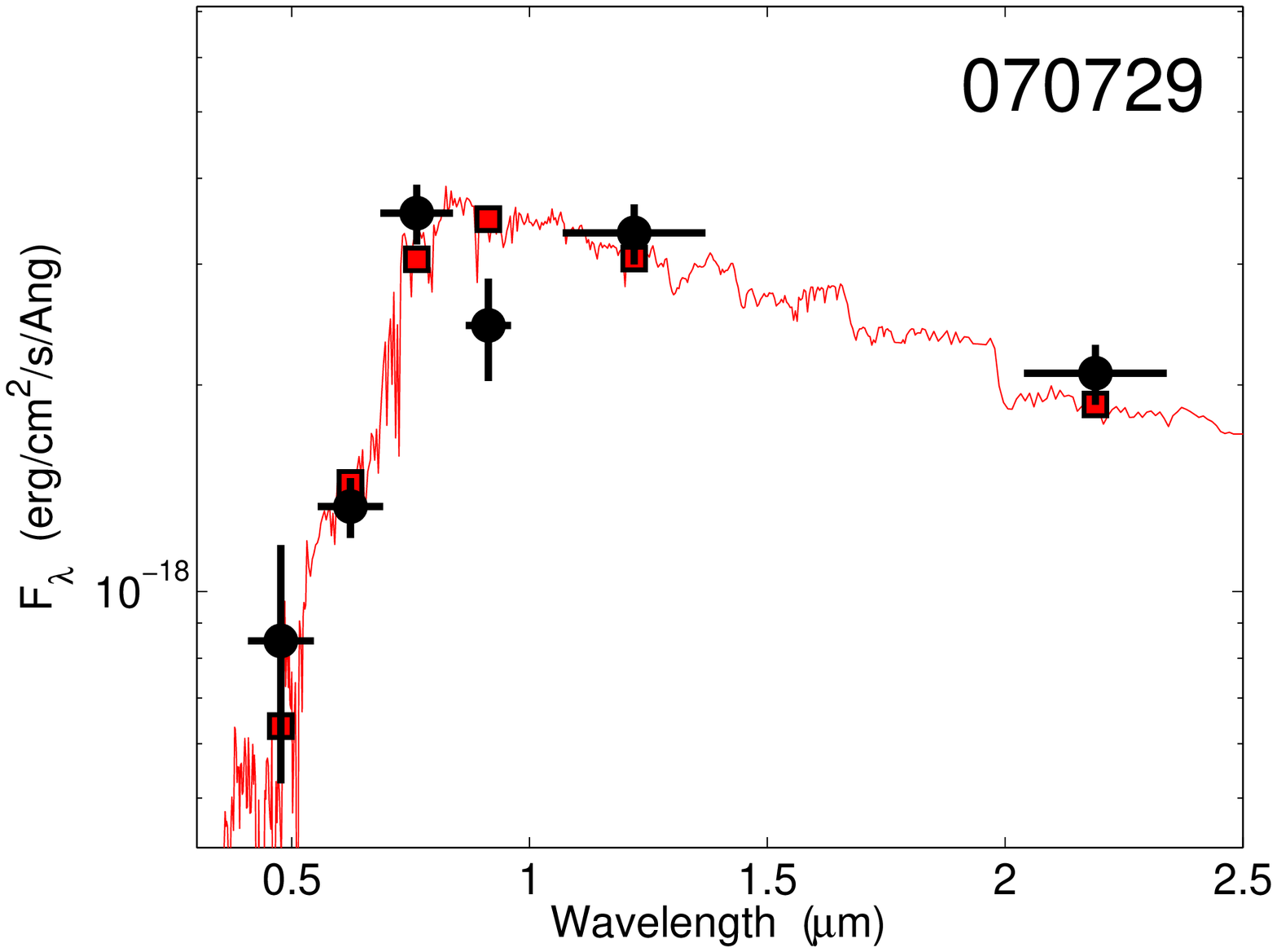}
\includegraphics[angle=0,width=1.725in]{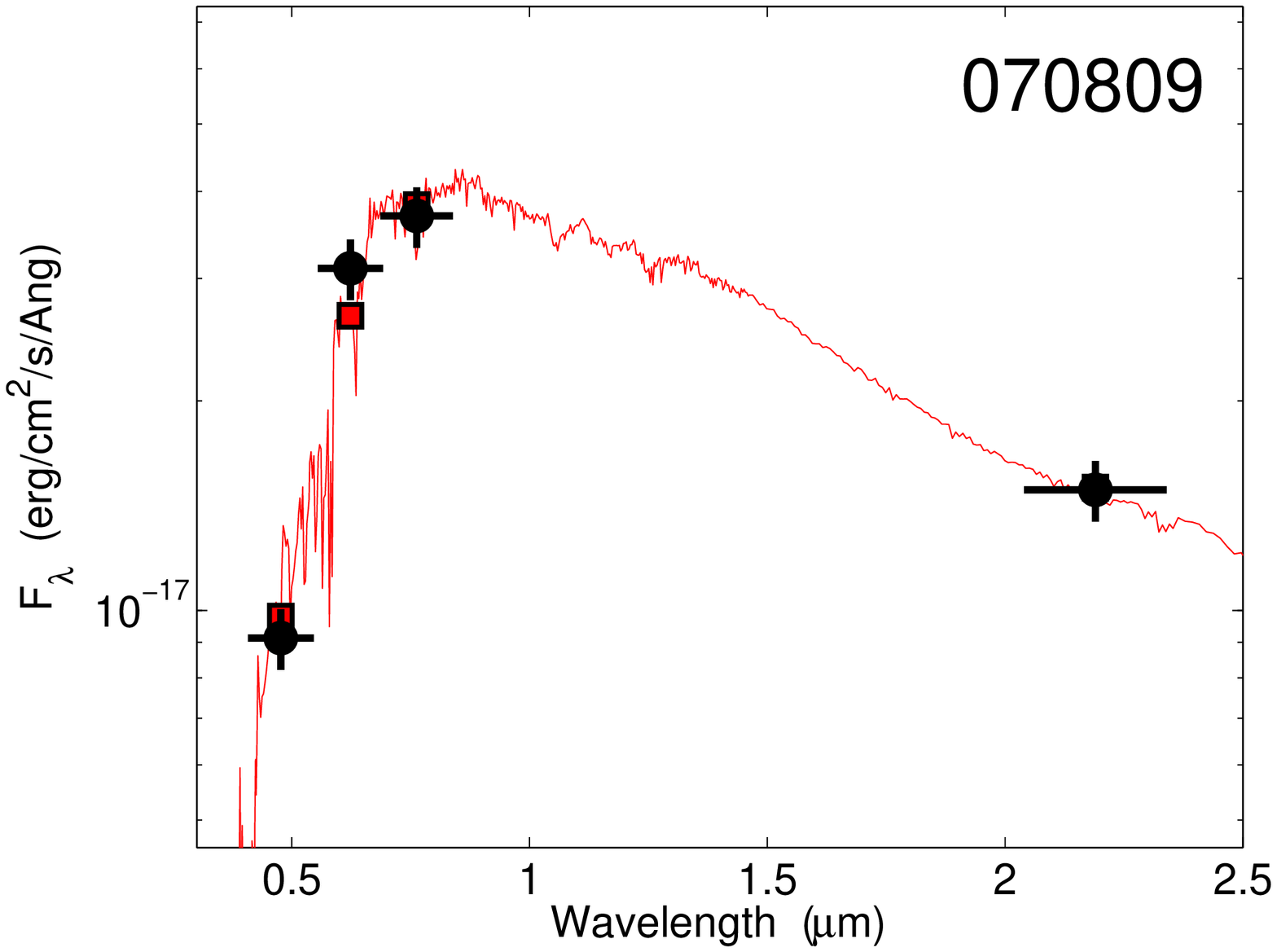}
\includegraphics[angle=0,width=1.725in]{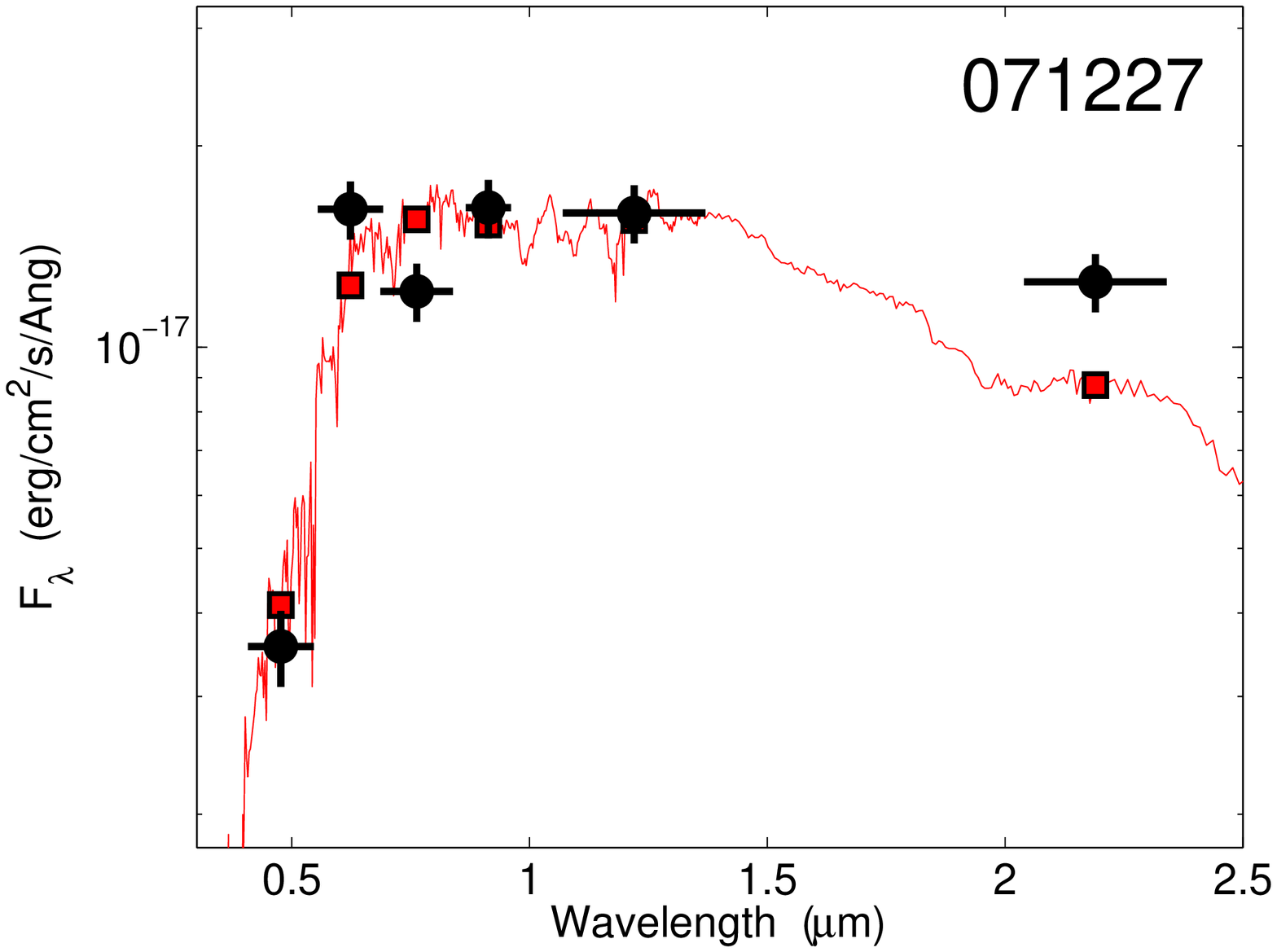}
\includegraphics[angle=0,width=1.725in]{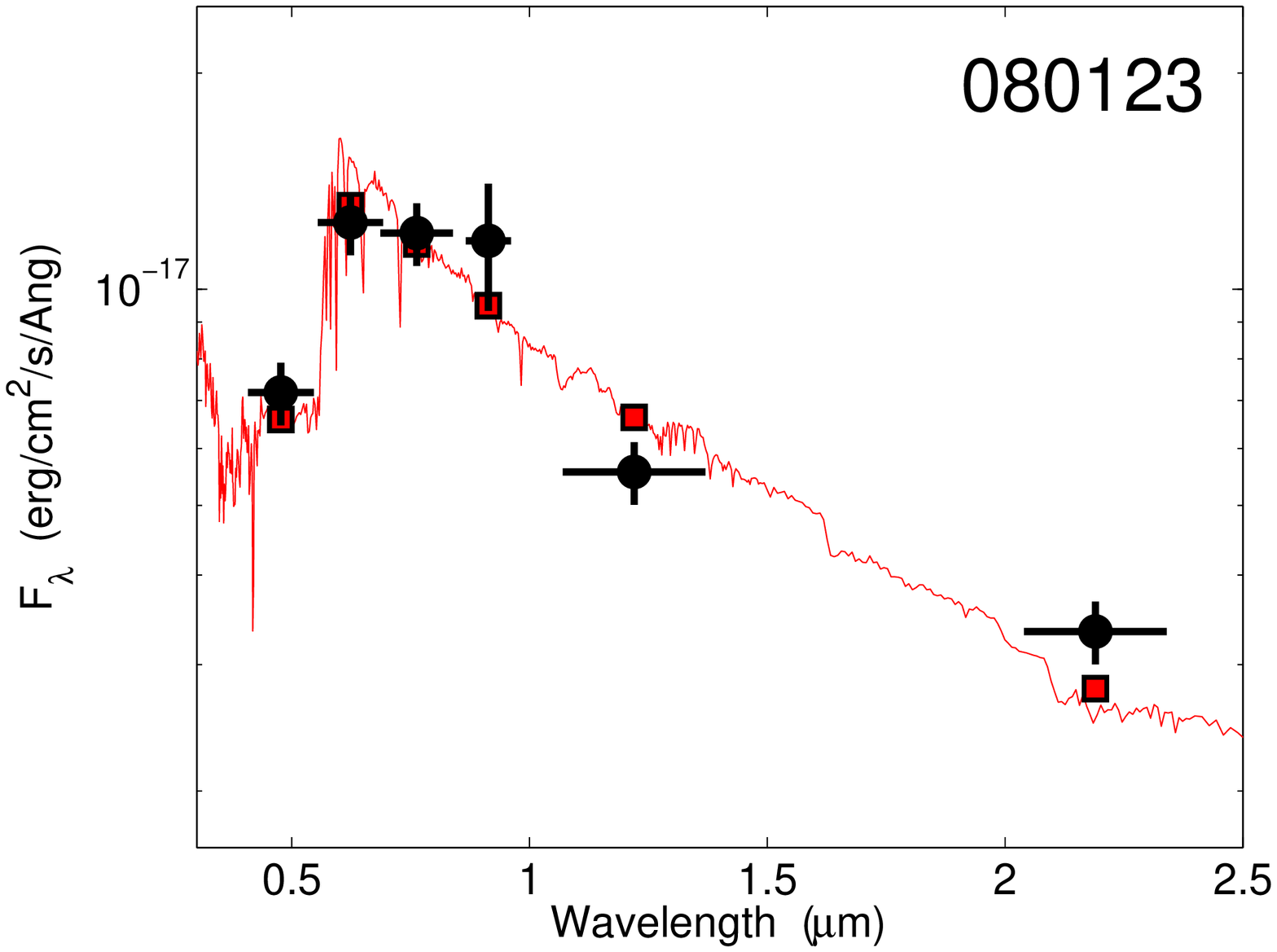}\\
\includegraphics[angle=0,width=1.725in]{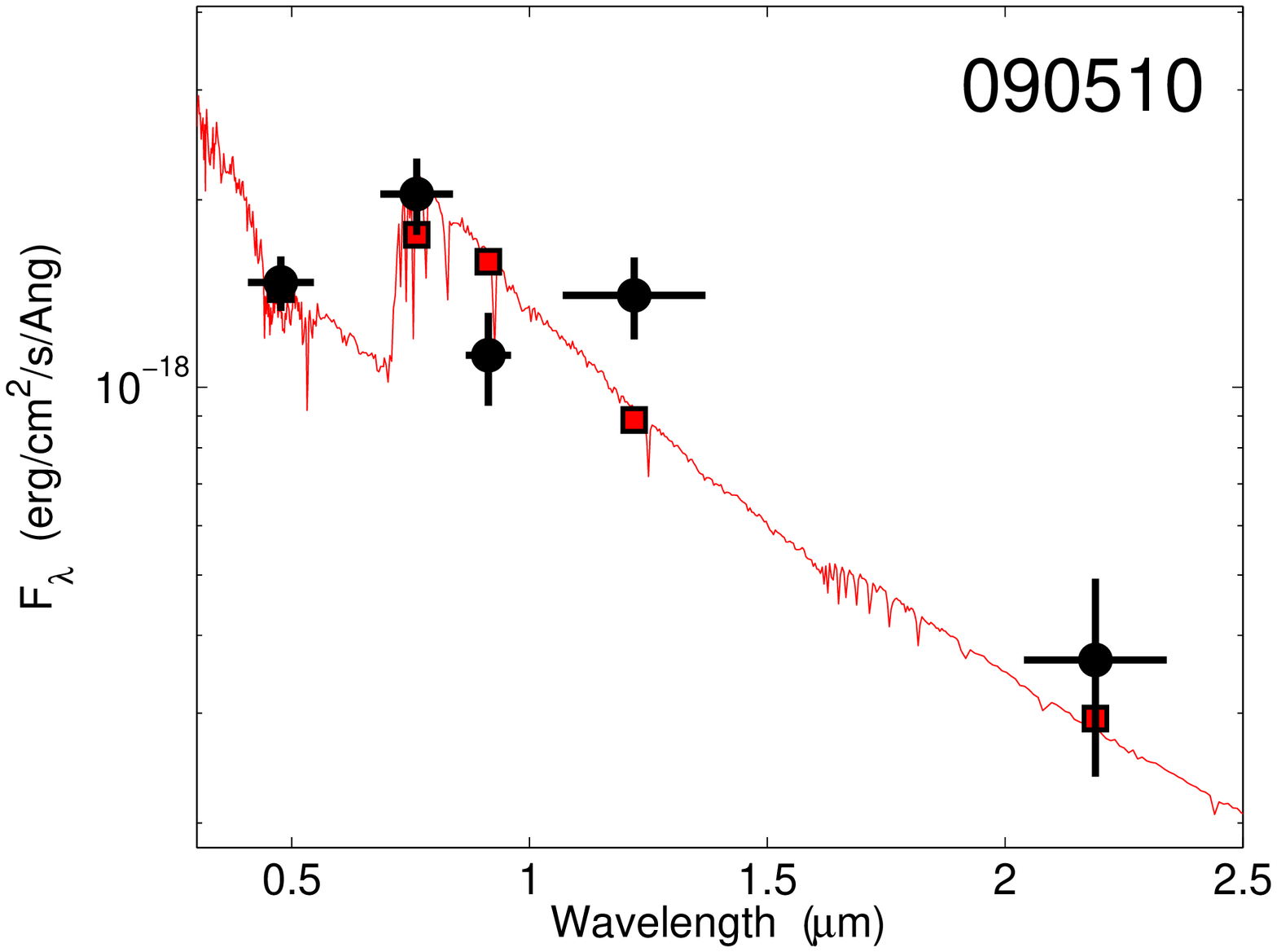}
\includegraphics[angle=0,width=1.725in]{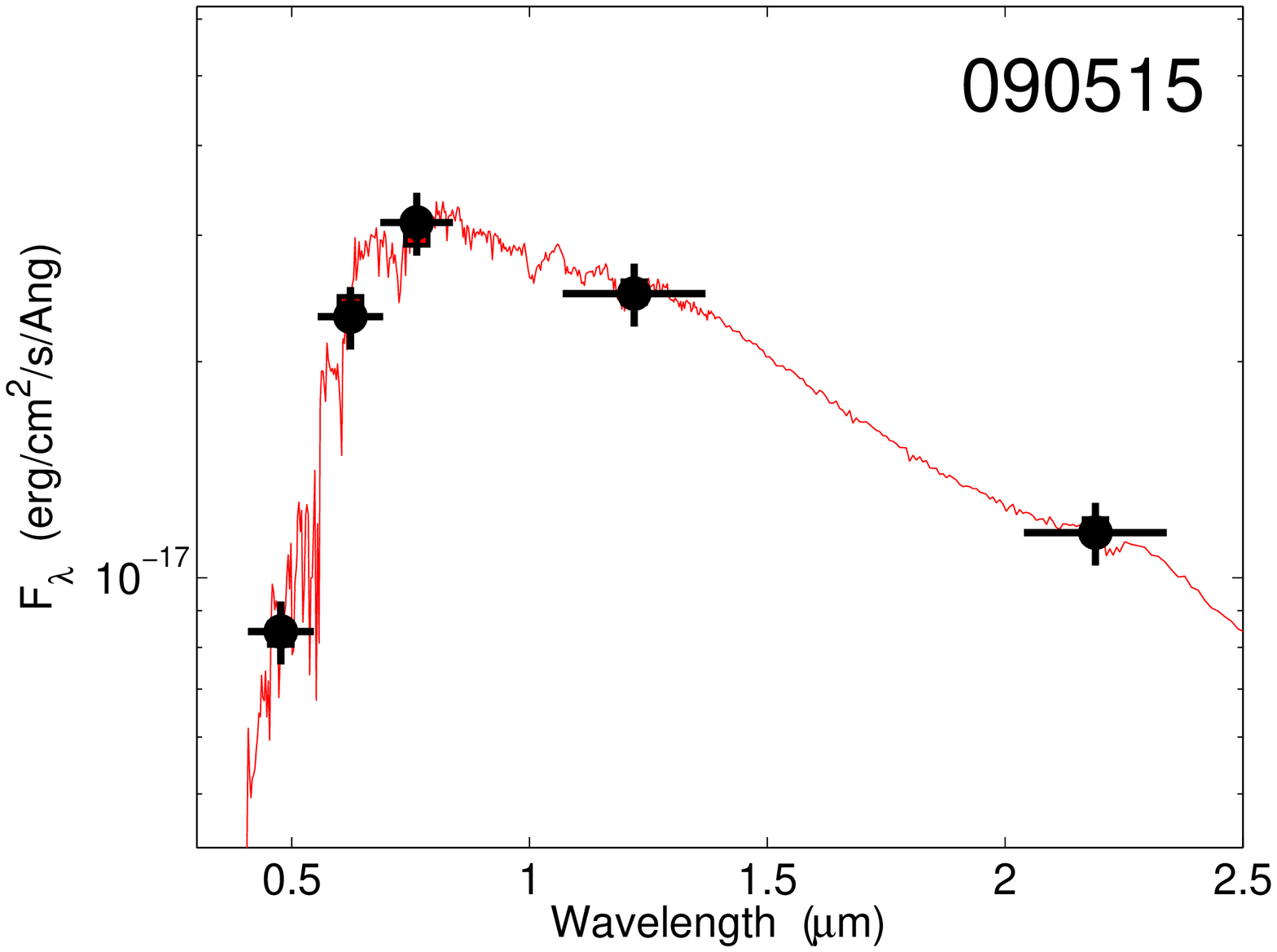}
\includegraphics[angle=0,width=1.725in]{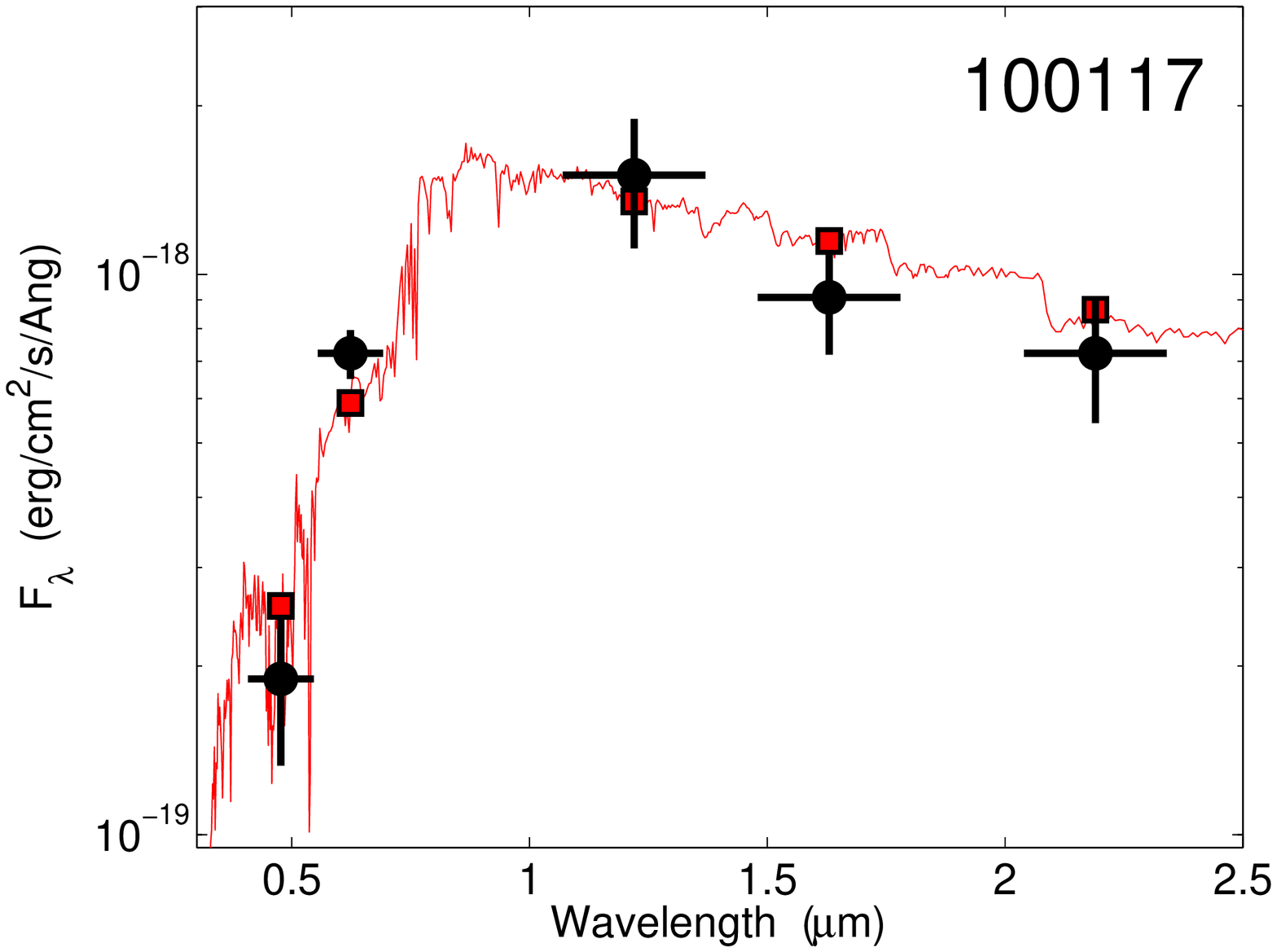}
\caption{Optical and near-IR spectral energy distributions of short
GRB host galaxies (black circles).  Each SED is fitted with a
\citet{mar05} single stellar population model (red line) through a
maximum likelihood fit of the synthesized photometry (red squares).
The resulting best-fit mass and age are listed in
Table~\ref{tab:short}.
\label{fig:seds}} 
\end{figure}

Reduction and photometry of the Magellan observations were performed
using standard routines in IRAF, including dark frame subtraction and
fringe correction for the PANIC data.  The Gemini data were processed
using the {\tt gemini} package in IRAF.  The resulting photometric
measurements (in AB magnitudes and corrected for Galactic extinction
using \citealt{sfd98}) are listed in Table~\ref{tab:data} and the
spectral energy distributions (SEDs) are shown in
Figure~\ref{fig:seds}.  As noted in Table~\ref{tab:data}, our host
observations are supplemented by data from the literature for GRBs
050709 \citep{hwf+05,cmi+06}, 050724 \citep{gcg+06}, 051221
\citep{sbk+06}, 061006 \citep{dmc+09}, and 070714b \citep{gfl+09}; for
the host of GRB\,050509b we used photometric measurements from the
SDSS and 2MASS catalogs.

To convert fluxes to luminosities we use the standard cosmological
parameters: $H_0=70$ km s$^{-1}$ Mpc$^{-1}$, $\Omega_m=0.27$, and
$\Omega_\Lambda=0.73$.

\section{Stellar Population Modeling}
\label{sec:model}

To determine the stellar masses and population ages of the short GRB
host galaxies we fit the optical/near-IR SEDs with the stellar
population models of \citet{mar05}.  These models provide the spectral
luminosity as a function of wavelength for an equivalent of 1
M$_\odot$.  We use the single age simple stellar population (SSP) red
horizontal branch morphology models with a Salpeter initial mass
function.  We further use Solar metallicity as generally appropriate
for short GRB hosts \citep{ber09}; we note that changing the
metallicity to $0.5$ Z$_\odot$ results in a systematic decrease in the
inferred masses and ages by at most $\sim 0.1$ dex.

To fit the observed SEDs with the SSP models, which exist for a
discrete set of ages, we create an interpolated grid with several
hundred logarithmically-spaced ages between 1 Myr and the age of the
universe at each host's redshift.  We then transform the model
spectral luminosities and wavelengths to the observer frame and
convolve with the appropriate filter response functions to create
model magnitudes.  The best-fit model for each host is determined
using $\chi^2$ minimization with the age of the population as the free
parameter and the best-fit mass for each age (i.e., the normalization)
determined by the equation:
\begin{equation}
M_{\rm bf}=\frac{\sum_{i=1}^{N}{F_{\rm\nu ,model,i}\times F_{\rm\nu
,data,i}/ \sigma_{\rm\nu ,data,i}^2}}{\sum_{i=1}^{N}{F_{\rm\nu
,model,i}^2/\sigma_{\rm\nu ,data,i}^2}},
\label{eqn:mass} 
\end{equation}
where $F_{\rm\nu ,model,i}$ are the model fluxes, $F_{\rm\nu ,data,i}$
and $\sigma_{\rm\nu ,data,i}$ are the observed fluxes and
uncertainties, respectively (Table~\ref{tab:data}), and $N$ is the
number of data points.  The resulting best-fit single age SSP models
are shown in Figure~\ref{fig:seds}, and the inferred masses and ages
are listed in Table~\ref{tab:short}.

The SSP models provide a robust estimate of the stellar masses and
population ages for the early-type host galaxies, since they are
dominated by old stellar populations, which exhibit only a mild
variation in the mass-to-light ratio as a function of age.  However,
for the late-type hosts, which have on-going star formation, the SSP
fits essentially provide light-weighted (rather than mass-weighted)
values, and they can therefore severely under-estimate the total
stellar mass and the mass-weighted age of the stellar population.
These are the quantities we are interested in for a determination of
whether short GRBs track stellar mass.  We therefore use two
approaches to assess the impact of older stellar populations for these
hosts.

\begin{figure}
\centering
\includegraphics[angle=0,width=2.3in]{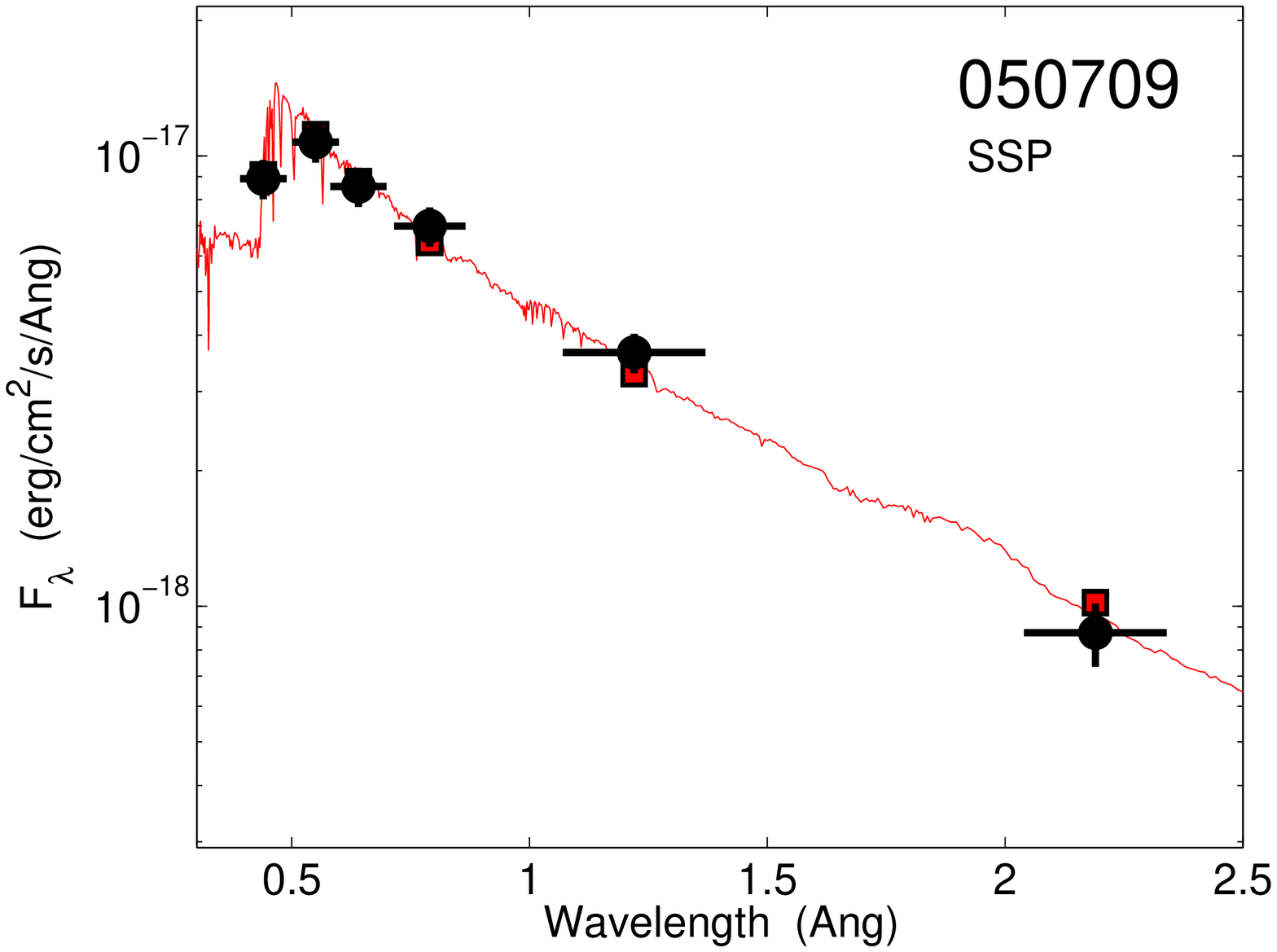}
\includegraphics[angle=0,width=2.3in]{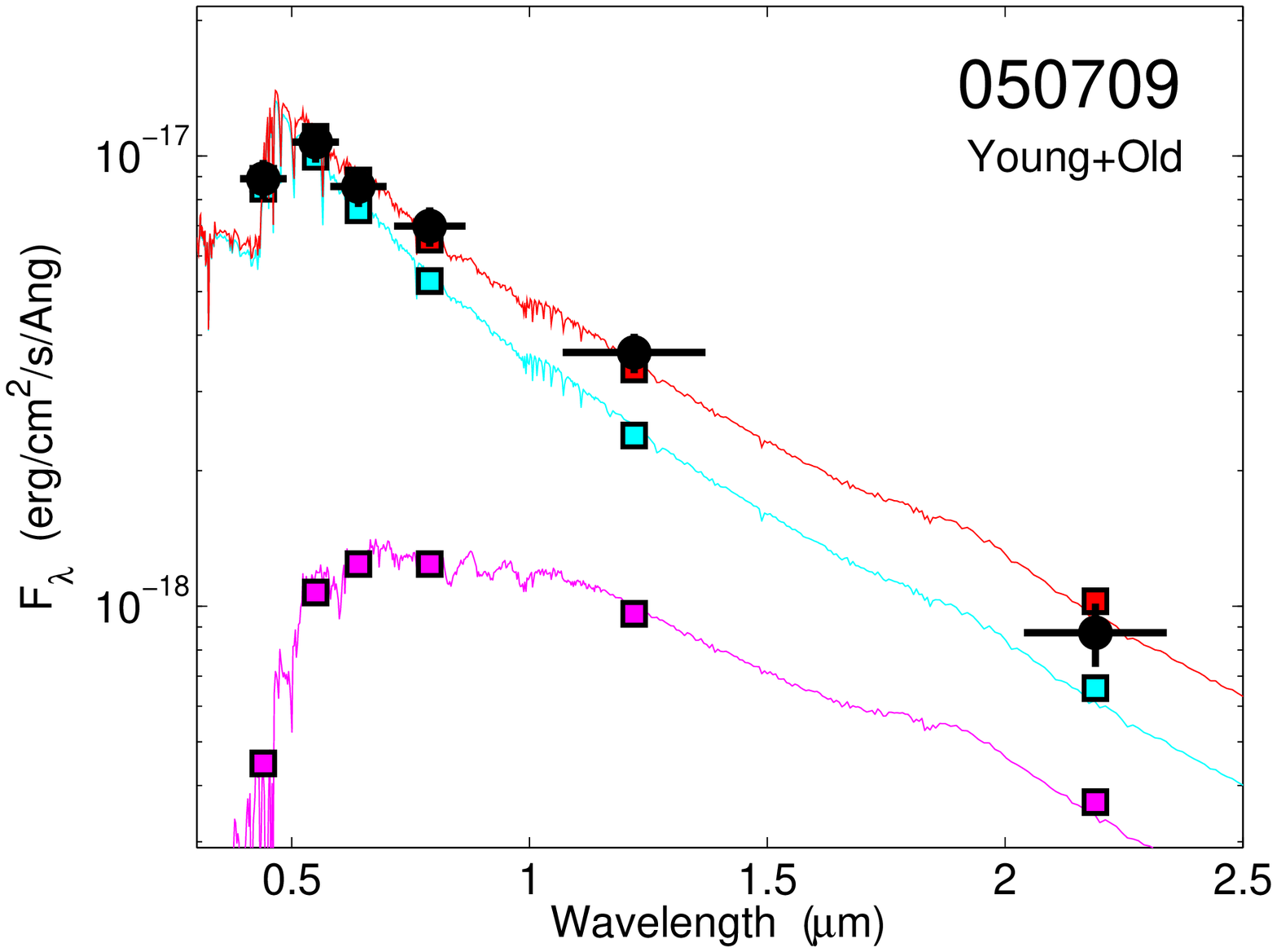}
\includegraphics[angle=0,width=2.3in]{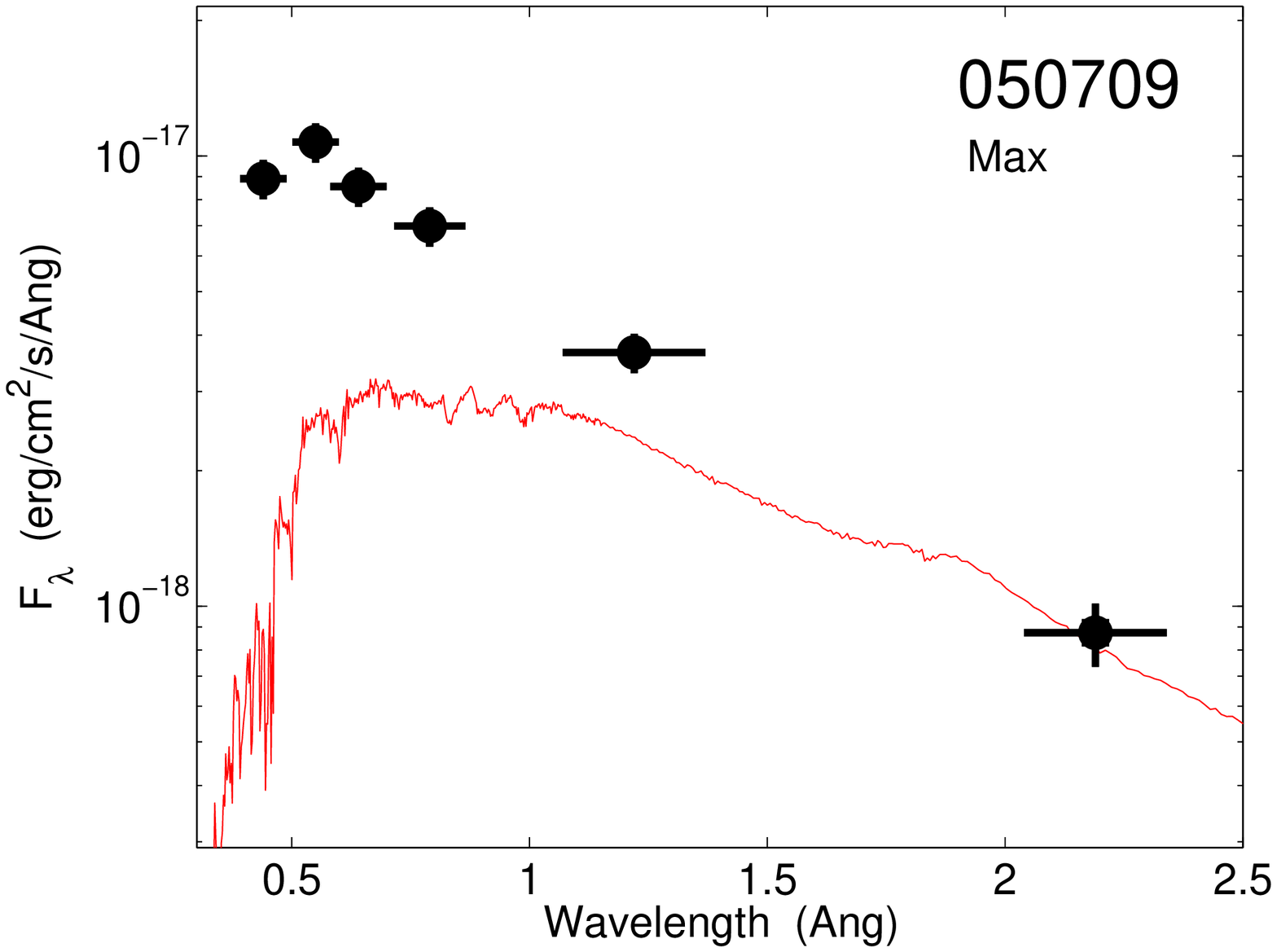}
\caption{Optical and near-IR spectral energy distribution of
GRB\,050709 with the three models used in this paper to extract the
stellar mass and population age.  Symbols are as in
Figure~\ref{fig:seds}.  {\it Left:} Single age SSP model.  {\it
Center:} Young+Old SSP model (magenta=old; cyan=young) with the old
population age fixed at the age of the universe at the redshift of the
burst ($z=0.161$ in this case).  {\it Right:} Maximal mass model with
the population age fixed at the age of the universe and using only the
$K$-band photometry.  The Young+Old model leads to total masses
intermediate between the single age SSP and the maximal models, and
has younger ages for the young population than the single age SSP
model.  The resulting best-fit masses and ages for the three models
are listed in Table~\ref{tab:short}.
\label{fig:seds2}} 
\end{figure}

First, we determine the maximum possible masses by assuming that the
mass of each host galaxy is dominated by a stellar population with the
age of the universe at its redshift.  Using the inferred (maximal)
mass-to-light ratios from the \citet{mar05} models we use the
available $K$-band fluxes to extract the maximal masses.  The
resulting values are listed in Table~\ref{tab:short}, and provide an
upper bound on the mass of each galaxy.  An example of this approach
compared to the single age SSP model is shown in
Figure~\ref{fig:seds2}.

Second, to provide a more realistic estimate we use a combined
young+old SSP model to fit the entire SED of each late-type host
galaxy and hence to determine the breakdown of masses in the young and
old populations.  The old population is fixed at the age of the
universe at each host redshift, and we fit for the age of the young
population and the mass ratio of the two populations; the old
population mass is given by Equation~\ref{eqn:mass} and the young
population mass then trivially follows from the best-fit ratio.  The
results are summarized in Table~\ref{tab:short} and an example of this
approach compared to the single age SSP model is shown in
Figure~\ref{fig:seds2}.  We note that for some of the hosts a wide
range of solutions is possible, including negligible mass in the old
population (i.e., a solution identical to the original single age SSP
fit).  For these hosts we list the maximum possible mass in the old
component within $1\sigma$ of the best fit.  As expected, the
young+old models lead to total masses intermediate between the single
age SSP and the maximal masses.  The addition of the old population
also leads to systematically younger ages for the young population
relative to the best-fit single age SSP values.

With the exception of the host of GRB\,071227 we find that host
extinction is not required for a satisfactory fit of the observed
SEDs.  The addition of host extinction will generally result in
systematically younger ages and somewhat higher masses for the young
stellar population due to the increase in UV luminosity.  On the other
hand, extinction does not significantly affect the total stellar mass
since it is determined by the near-IR luminosity.  We find that the
addition of $A_V\approx 0.5$ mag in the rest-frame of the host tends
to change the SSP ages and masses (or the young population ages and
masses in the young+old models) by at most $-0.3$ dex and $+0.1$ dex,
respectively.  Larger extinction values significantly degrade the
goodness of fit.

Finally, to uniformly compare the stellar masses and ages of short and
long GRB hosts, we fit the SEDs of the long GRB hosts from
\citet{sgl09} in a similar redshift range ($z\approx 0.03-1.6$) with
the same stellar population models and techniques used here, but with
a sub-solar metallicity.  We derive single age SSP values, and maximal
masses using the $K$-band photometry alone.  The results are listed in
Table~\ref{tab:long}.  As in the case of the short GRB hosts, we do
not include the effects of extinction, unless strictly required by the
goodness of fit (see Table~\ref{tab:long}).

\section{Host Galaxy Stellar Masses and Ages}
\label{sec:res}

\begin{figure}
\epsscale{1}
\plotone{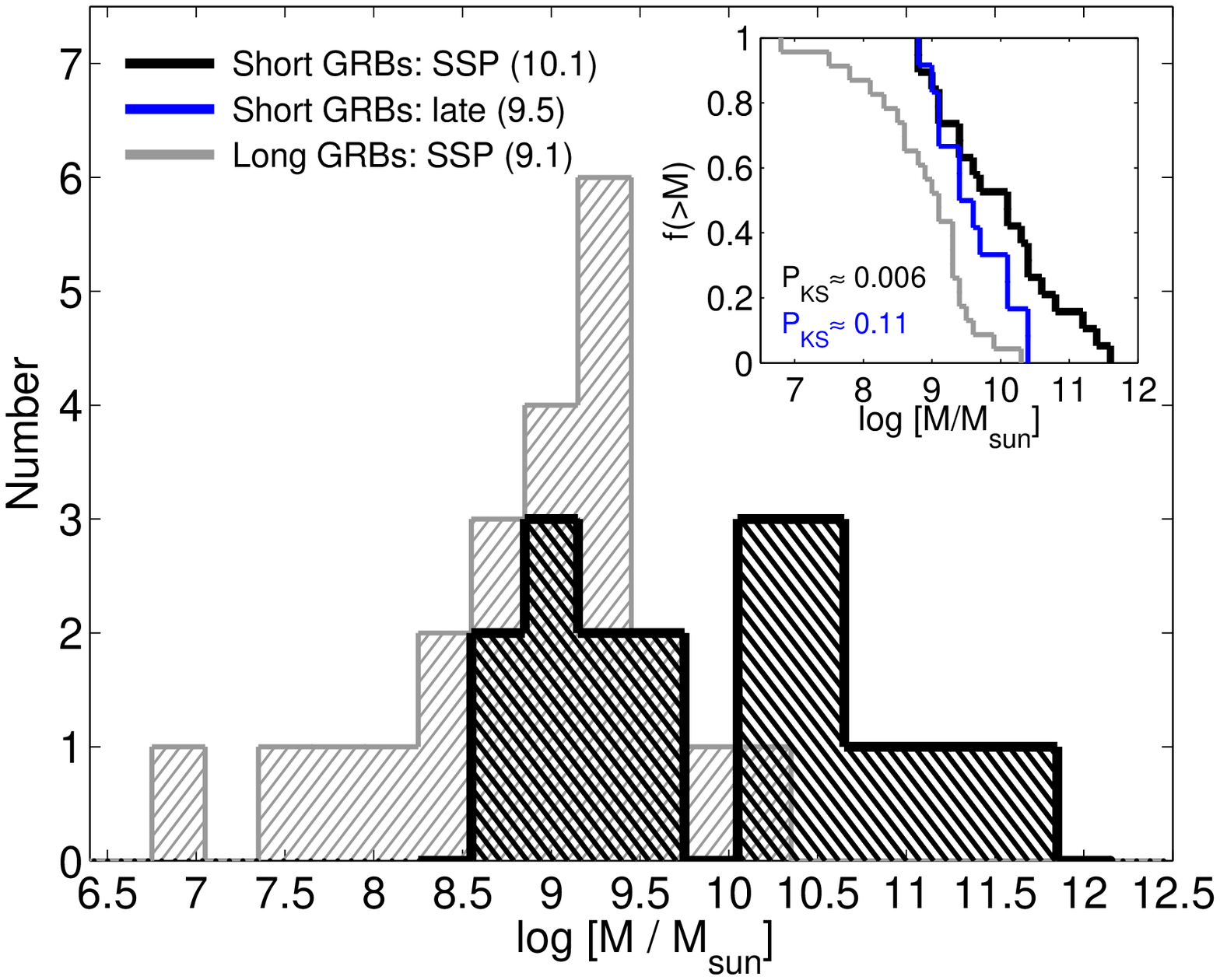}
\caption{Histograms of inferred stellar masses from the single stellar
population fits shown in Figure~\ref{fig:seds} for the hosts of short
(black) and long (gray) GRBs.  The inset shows the cumulative
distributions, including for the subset of late-type short GRB hosts
(blue).  The median values for the three samples are given in
parentheses, and the Kolmogorov-Smirnov probabilities that the
distributions of short and long GRB hosts, as well as star forming
short GRB and long GRB hosts are drawn from the same distribution are
provided in the inset.
\label{fig:masses_1a}} 
\end{figure}

The distribution of stellar masses derived from the single age SSP
fits is shown in Figure~\ref{fig:masses_1a}.  We find that the masses
span three orders of magnitude, $M_{\rm SSP}\approx 6\times
10^8-4\times 10^{11}$ M$_\odot$, with a median value of $\langle
M_{\rm SSP}\rangle\approx 1.3\times 10^{10}$ M$_\odot$.  Dividing the
sample into early- and late-type host galaxies we find that the former
span the range $M_{\rm SSP}\approx (2-40)\times 10^{10}$ M$_\odot$,
while the latter have much lower masses of $M_{\rm SSP}\approx
(0.06-2)\times 10^{10}$ M$_\odot$.  The clear distinction between the
two samples partially reflects the bias of single age SSP models,
which for the late-type hosts are dominated by the young stellar
population and hence under-estimate the contribution of any older
stellar populations.  If we restrict the sample to the host galaxies
identified in coincidence with optical afterglows we find a median
mass of $\langle M_{\rm SSP}\rangle\approx 5\times 10^{9}$ M$_\odot$.
This value is lower than for the full sample because 3 of the 5
early-type hosts, which have the largest SSP masses, are associated
with XRT positions (GRB\,050509b) or inferred from chance coincidence
arguments (GRBs 070809 and 090515).

\begin{figure}
\epsscale{1}
\plotone{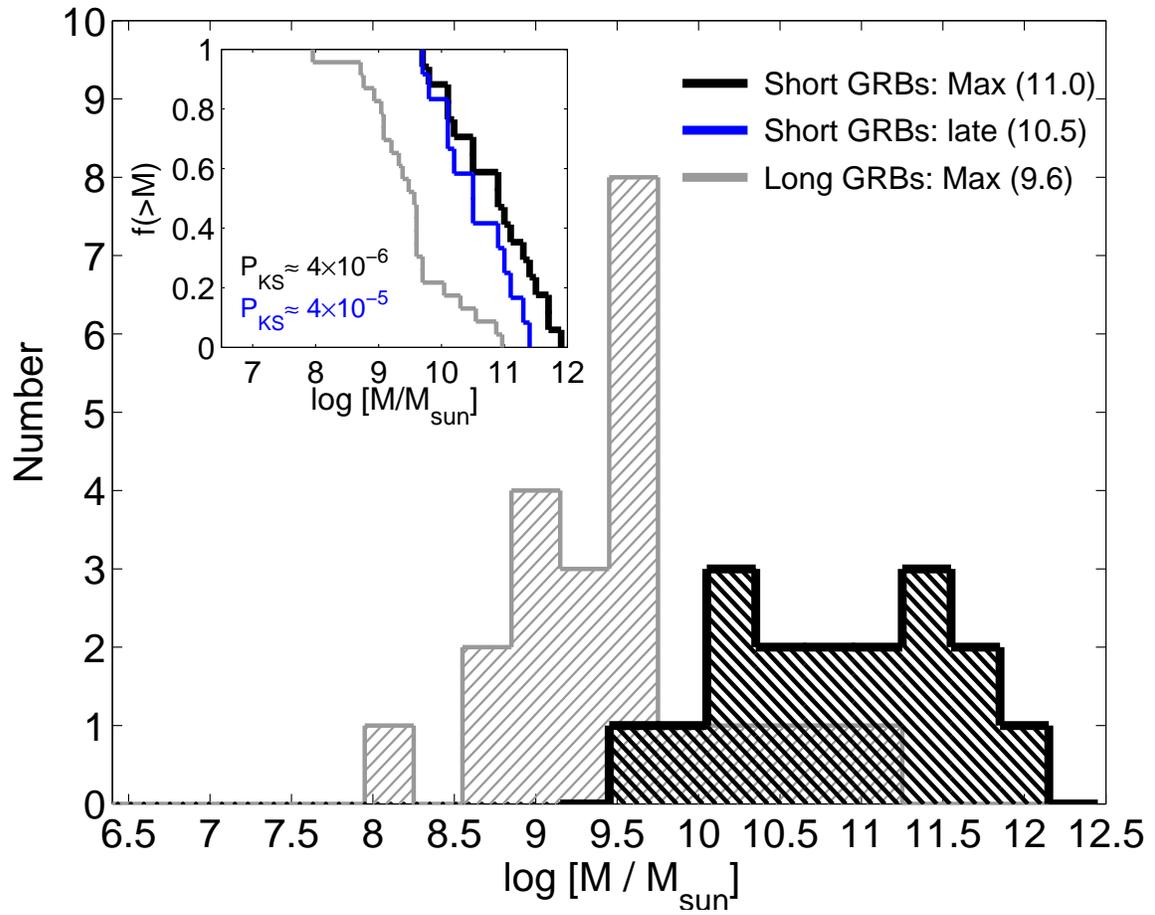}
\caption{Same as Figure~\ref{fig:masses_1a} but for the maximal
masses.
\label{fig:masses_1b}} 
\end{figure}

The differences between the masses of the early- and late-type hosts
are less severe in the case of the maximal masses
(Figure~\ref{fig:masses_1b}).  For the population as a whole we find a
reduced range of $M_{\rm Max}\approx 6\times 10^9-8\times 10^{11}$
M$_\odot$.  The median mass is $\langle M_{\rm Max}\rangle\approx
1\times 10^{11}$ M$_\odot$, about an order of magnitude larger than
for the single age SSP masses, and only slightly larger than the
stellar mass of the Milky Way.  As expected, the ratios of maximal to
SSP masses for the early-type hosts are modest, $M_{\rm Max}/M_{\rm
SSP}\approx 2-8$, since these hosts are already dominated by old
stellar populations.  However, for the late-type hosts the corrections
are significant, $M_{\rm Max}/M_{\rm SSP}\approx 5-60$, with a median
ratio of about an order of magnitude.  If we again restrict the sample
to host galaxies identified in coincidence with optical afterglows we
find a median mass of $\langle M_{\rm Max}\rangle\approx 3\times
10^{10}$ M$_\odot$, lower than for the full sample due to the
preferential rejection of early-type hosts.

While the maximal masses provide a robust upper bound, they are likely
to over-estimate the true stellar mass, particularly for the
early-type hosts which do not suffer from the young-population bias
present in the late-type hosts.  If we instead use the young+old
models for the late-type hosts, and the single age SSP values for the
early-type hosts, we find masses of $M\approx 2\times 10^9-4\times
10^{11}$ M$_\odot$, with a median of $\langle M\rangle\approx 5\times
10^{10}$ M$_\odot$.  The cumulative distributions of stellar masses
for the host population as a whole, and for the subsets of early- and
late-type galaxies are shown in Figure~\ref{fig:masses_2}.  The Figure
clearly highlights the larger masses of the early-type hosts; even
when comparing SSP masses for the early-type hosts with maximal masses
for the late-type hosts the median for the former is larger by about a
factor of 5.

\begin{figure}
\epsscale{1}
\plotone{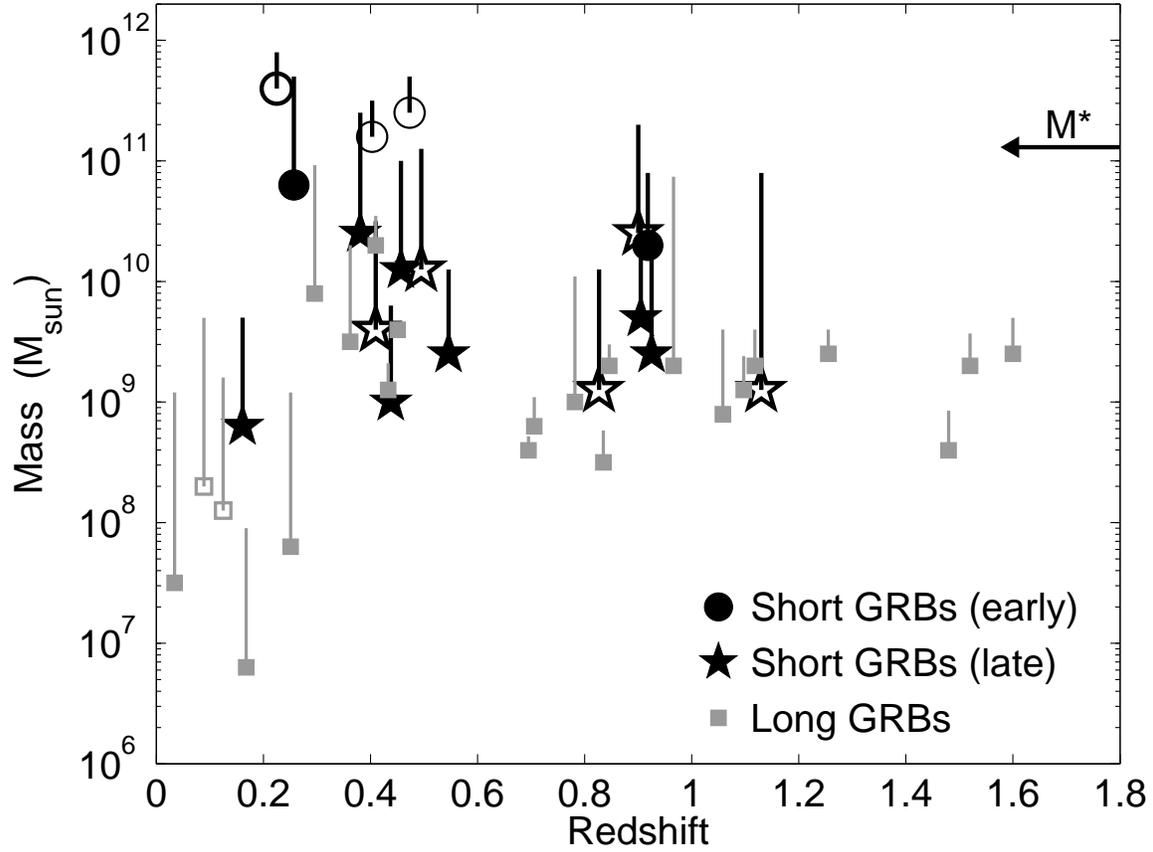}
\caption{Single stellar population masses plotted as function of
redshift for the hosts of short (black) and long (gray) GRBs.  Circles
and stars designate early- and late-type short GRB hosts,
respectively.  Filled symbols designate hosts identified through
optical afterglow positions, thick open symbols designate hosts
identified through coincidence with {\it Swift}/XRT error circles, and
thin open symbols designate galaxies identified as potential hosts for
short GRBs with optical afterglows based on chance coincidence
probabilities \citep{ber10}.  Open gray symbols designate the
ambiguous long GRBs 060505 and 060614.  In all cases the lines
indicate the maximal masses.
\label{fig:mass_z}} 
\end{figure}

The range of inferred stellar masses does not appear to depend on
redshift.  As shown in Figure~\ref{fig:mass_z}, the stellar masses are
nearly uniform with redshift, both for the single age SSP and maximal
values.  We do find a preference for lower redshift among the subset
of early-type hosts, and since these galaxies are more massive, they
lead to somewhat higher masses at lower redshift ($z\sim 0.4$
vs.~$z\sim 0.9$).  A similar conclusion is reached even if we
restrict the sample to only the hosts identified in coincidence with
optical afterglows.

\begin{figure}
\epsscale{1}
\plotone{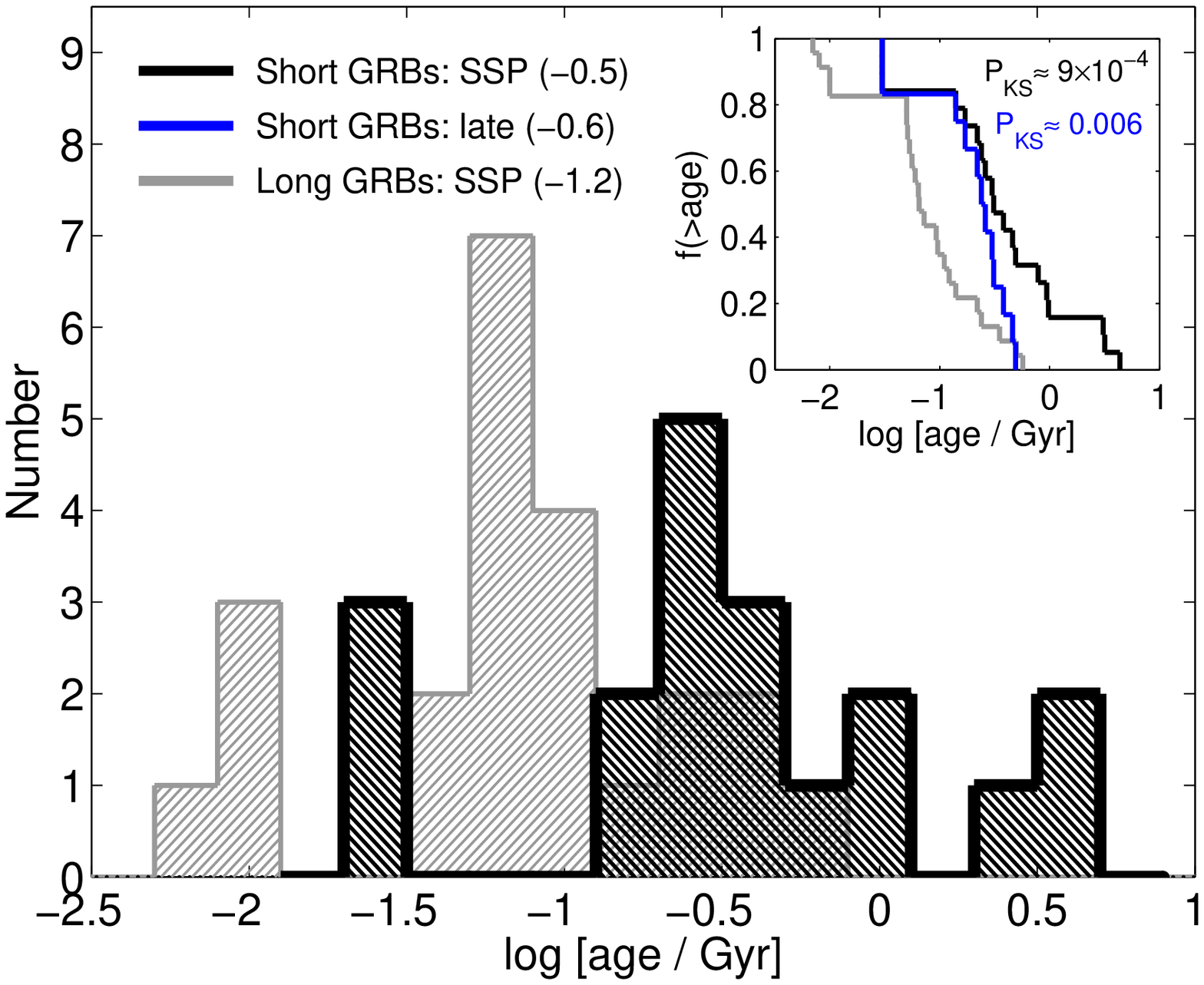}
\caption{Histograms of inferred stellar population ages from the
single stellar population fits shown in Figure~\ref{fig:seds} for the
hosts of short (black) and long (gray) GRBs.  The inset shows the
cumulative distributions, including for the subset of late-type short
GRB hosts (blue).  The median values for the three samples are given
in parentheses, and the Kolmogorov-Smirnov probabilities that the
distributions of short and long GRB hosts, as well as star forming
short GRB and long GRB hosts are drawn from the same distribution are
provided in the inset.
\label{fig:ages}} 
\end{figure}

We next turn to the distribution of stellar population ages.  These
values are only available for the single age SSP models since for the
maximal and young+old models we assume a population with the age of
the universe at each host redshift.  The distribution of ages is shown
in Figure~\ref{fig:ages}, with the values ranging from about 30 Myr to
4.4 Gyr.  The median age is $\langle\tau_{\rm SSP}\rangle \approx 0.3$
Gyr for the full sample, with $\langle\tau_{\rm SSP}\rangle\approx
0.25$ Gyr for the subset of late-type hosts and $\langle\tau_{\rm
SSP}\rangle\approx 3$ Gyr for the subset of early-type hosts.  The
median value for the subset of hosts identified in coincidence with
optical afterglows is identical to the median for the late-type hosts.
As indicated above, the young+old models result in somewhat younger
ages for the young stellar population since some of the near-IR flux
is accounted for by the forced old population.  The median age for the
young population in this model is about 0.1 Gyr
(Table~\ref{tab:short}).

\begin{figure}
\epsscale{1}
\plotone{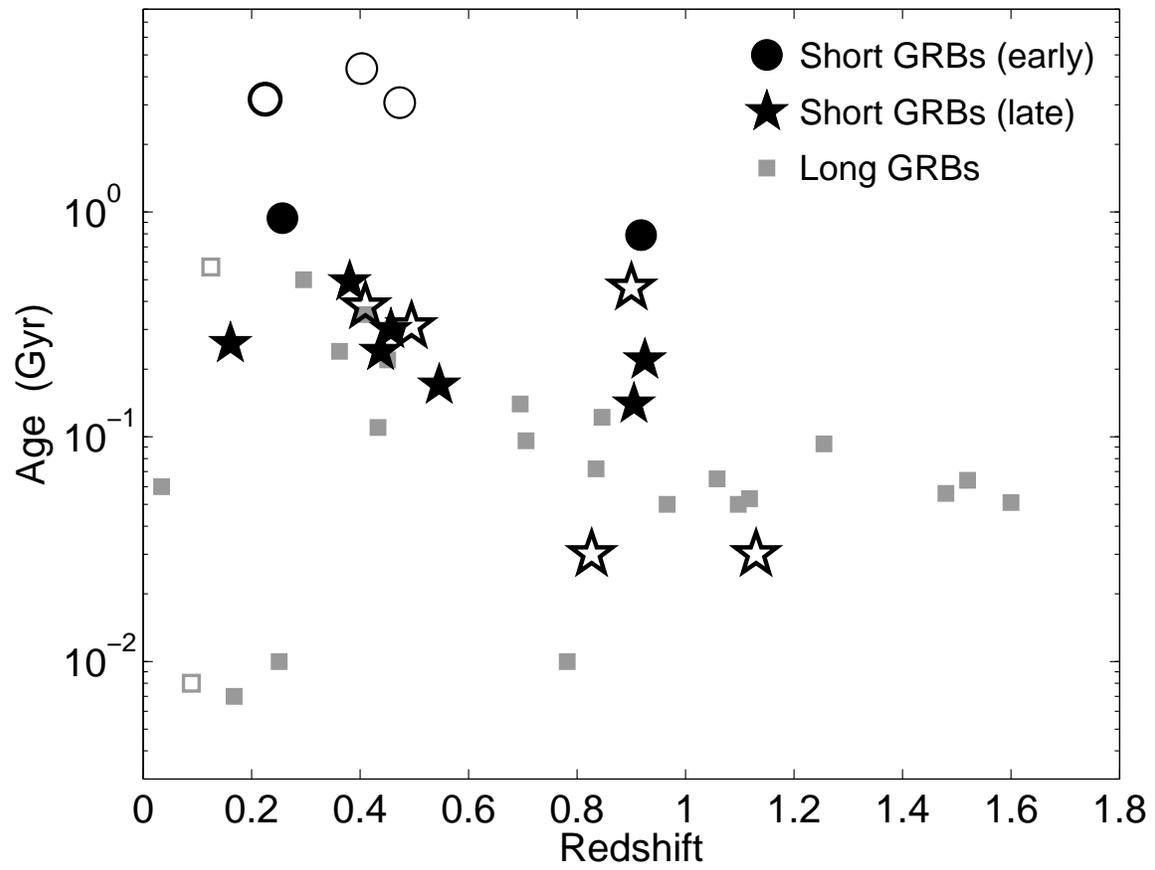}
\caption{Single stellar population ages plotted as function of
redshift for the hosts of short (black) and long (gray) GRBs.  Symbols
are as in Figure~\ref{fig:mass_z}.
\label{fig:ages_z}} 
\end{figure}

Finally, in Figure~\ref{fig:ages_z} we plot age as a function of
redshift for the single age SSP models.  For the full sample we find a
significant negative correlation between the two quantities, with a
Kendall's $\tau$ value of $-0.57$, corresponding to a null hypothesis
(no-correlation) probability of only $\approx 0.01$.  However, the
correlation appears to be due mainly to hosts identified in
coincidence with XRT error circles and through chance coincidence
probabilities.  If we restrict the sample to hosts identified in
coincidence with optical afterglows, then no significant correlation
is found (Kendall's $\tau$ value of -0.33 with a no-correlation
probability of $0.26$).  We note that for the maximal and young+old
models we cannot associate an age with the old stellar population due
to our conservative assumption that it is equivalent to the age of the
universe.

\subsection{Long GRB Host Galaxies}

For the long GRB host comparison sample we derive single age SSP
parameters, as well as maximal masses.  For the SSP model we find a
range of $M_{\rm SSP}\approx 6\times 10^6-2\times 10^{10}$ M$_\odot$,
with a median value of $\langle M_{\rm SSP}\rangle\approx 1.3\times
10^{9}$ M$_\odot$ (Figure~\ref{fig:masses_1a}).  The maximal masses
span $M_{\rm Max}\approx 9\times 10^7-9\times 10^{10}$ M$_\odot$, with
a median value of $\langle M_{\rm Max}\rangle\approx 4.0\times 10^{9}$
M$_\odot$ (Figure~\ref{fig:masses_1b}).  Our median SSP and maximal
values bracket the median masses of $2\times 10^9$ M$_\odot$ and
$1.8\times 10^9$ M$_\odot$ found by \citet{sgl09} and \citet{lkb+10},
respectively.  We also note that the median ratio of $M_{\rm
Max}/M_{\rm SSP}\approx 3$ for the long GRB hosts is more modest than
for the late-type short GRB hosts (see above), indicating that long
GRB hosts do not generally harbor massive old stellar populations.

As in the case of the short GRB hosts, we find no clear correlation
between redshift and stellar mass, although the lowest redshift hosts
($z\lesssim 0.25$) appear to have lower than average masses
(Figure~\ref{fig:mass_z}).  These hosts also exhibit some of the
largest ratios of $M_{\rm Max}/M_{\rm SSP}$, so the low masses may be
at least partially due to their very young ages (4 of the 5 hosts are
consistent with ages of only $\sim 10$ Myr; see Table~\ref{tab:long}
and Figure~\ref{fig:ages_z}).

The SSP stellar population ages span about 10 to 570 Myr, with a
median value of $\langle \tau_{\rm SSP}\rangle\approx 65$ Myr.  We
find no clear correlation between the ages and redshifts of these
hosts, although there appears to be a larger spread in ages at
$z\lesssim 0.5$ than at higher redshifts.  The inferred ages are in
good agreement with the analysis of \citet{chg04} for a smaller
sample.

Finally, we pay particular attention to the hosts of GRBs 060505 and
060614 since these nearby bursts had long durations but lacked an
associated supernova, possibly pointing to a non-massive star origin
\citep{dcp+06,fwt+06,gfp+06,gnb+06,ocg+07,lk07,tfo+08}.  For the host
of GRB\,060505 we find a best-fit\footnotemark\footnotetext{This
includes a rest-frame extinction of $A_{\rm V,host}=0.3$ mag.  A model
with no extinction provides a poorer fit ($\chi^2_r=2.4$) and leads to
an older age of 0.23 Gyr and a larger mass of $2.5\times 10^9$
M$_\odot$.} SSP age of about 10 Myr, with a stellar mass of about
$2\times 10^8$ M$_\odot$, while for the host of GRB\,060614 we infer
an age of 0.57 Gyr and a mass of $1.3\times 10^8$ M$_\odot$
(Table~\ref{tab:long}).  The maximal masses are $5\times 10^9$
M$_\odot$ and $1.6\times 10^9$ M$_\odot$, respectively.  The inferred
masses exhibit better agreement with the distribution of long GRB
hosts than with short GRB hosts.  The very young stellar population
age for the host of GRB\,060505 is similarly reminiscent of the long
GRB host population (see also \citealt{ocg+07,tfo+08}), while the SSP
age for the host of GRB\,060614 is the largest of any of the long GRB
hosts in this paper, and more closely in line with the short GRB host
population.

\section{Discussion}
\label{sec:disc}

Having established the basic stellar population properties of the
short GRB hosts in our sample we now address the fundamental questions
listed in \S\ref{sec:intro} through a comparison to the general galaxy
mass function and the hosts of long GRBs.

\subsection{Comparison to the Mass Distribution of Galaxies}
\label{sec:mass}

The question of whether the short GRB rate tracks stellar mass alone
can be addressed through two statistical tests: (i) the fraction of
short GRBs in early- and late-type galaxies; and (ii) the relation of
the host stellar masses to the general galaxy mass function.

\subsubsection{Host Demographics}

In the redshift range under consideration here ($z\sim 0.2-1$) about
an equal fraction of the cosmic stellar mass budget resides in
early- and late-type galaxies (e.g., \citealt{isl+10}).  Therefore, if
short GRBs track stellar mass alone we expect a roughly one-to-one
ratio of galaxy types.  This does not appear to be the case.  For
example, within the sample of short GRBs with optical afterglows (19
events), only 2 are unambiguously hosted by early-type galaxies (GRBs
050724 and 100117; \citealt{bpc+05}, Fong et al. in prep.), while 8
are unambiguously hosted by late-type galaxies; the probability of
obtaining this ratio from an intrinsic one-to-one distribution is only
0.04.  The identity of the remaining 9 hosts is unclear at the present
due to their faintness or the lack of underlying galaxies at the burst
positions.  Still, unless nearly all of these bursts were hosted by
early-type galaxies, the resulting ratio appears to be skewed in favor
of late-type host galaxies with on-going star formation activity.  We
note that the same result holds true if we consider the bursts with
only X-ray afterglow positions and identified hosts.

Thus, the host galaxy demographics suggest that short GRBs do not
track stellar mass alone, or phrased alternatively, they do not have a
delay time distribution that is skewed to old ages of $\sim {\rm few}$
Gyr.  It is possible, however, that this result is influenced by
secondary factors such as the typical circumburst density or intrinsic
differences in the energy scale and afterglow brightness as a function
of galaxy type (possibly reminiscent of the differences in peak
luminosity for Type Ia supernovae in early- and late-type galaxies;
\citealt{htp+00,mdp06}).  If such differences lead to fainter
afterglows (or prompt emission) for short GRBs in early-type galaxies,
this would suppress the early-type fraction.  Although the modest size
of the host sample, and the substantial fraction of short GRBs with
only $\gamma$-ray positions ($\sim 1/3$ of all events), prevent
definitive conclusions, it does not appear that the optical afterglows
of short GRBs in early- and late-type galaxies are distinct
\citep{ber10}.

Thus, our preliminary conclusion from the host demographics is that
short GRBs do not appear to track the cosmic stellar mass fraction in
early- and late-type galaxies.

\subsubsection{Comparison to the Galaxy Mass Function}

\begin{figure}
\epsscale{1}
\plotone{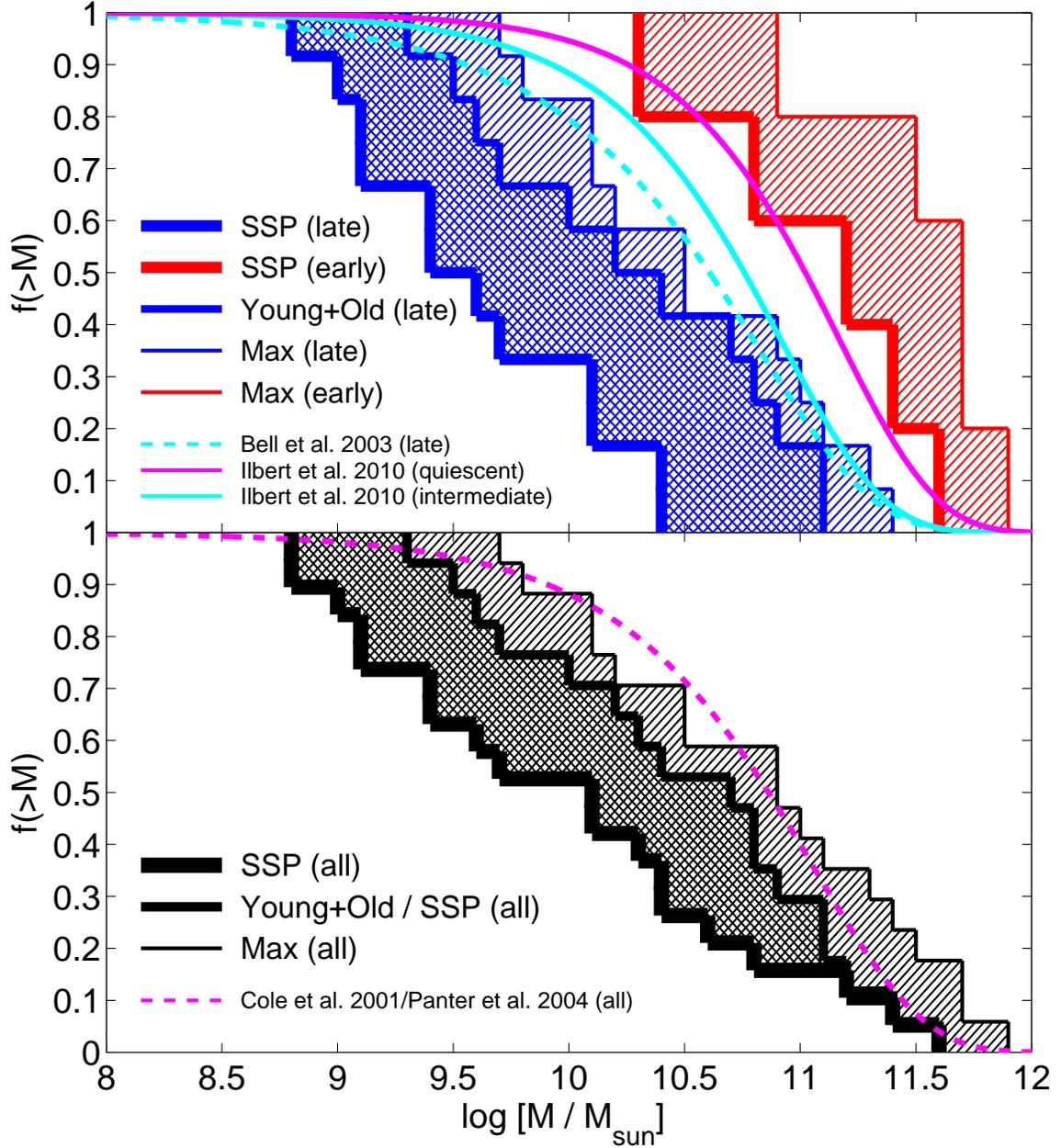}
\caption{Cumulative distributions for the full sample of single age
simple stellar population (SSP) masses, maximal masses, and combined
young+old and SSP masses for the late- and early-type hosts,
respectively (black; bottom panel).  The upper panel shows a breakdown
by galaxy type (late-type: blue; early-type: red).  The shaded regions
represent the range of possible stellar masses since the SSP masses,
which are effectively light-weighted values, are most likely an
under-estimate, while the maximal masses make the extreme assumption
that all hosts are dominated by populations with the age of the
universe.  For the late-type hosts we also plot the total masses from
a young+old SSP fit (Table~\ref{tab:short}), which are more closely
representative of the total mass.  Also shown are the fractions of
total stellar mass in galaxies with mass, $>M$, calculated from
several published galaxy stellar mass functions at $z\sim 0-2$ (cyan
and magenta lines; \citealt{cnb+01,bmk+03,phj04,isl+10}); for the
\citet{isl+10} mass function we use the $z\sim 0.5$ bin, appropriate
for the short GRB sample (Table~\ref{tab:data}), and separately plot
the mass function for quiescent galaxies and for intermediate-activity
galaxies, which resemble the intermediate star formation activity in
short GRB hosts \citep{ber09}.  The comparison indicates that short
GRBs trace galaxy mass {\it only} if the bulk of the late-type hosts
have close to maximal masses.  The subset of early-type hosts appears
to faithfully trace the mass function of galaxies for the SSP-derived
masses.
\label{fig:masses_2}} 
\end{figure}

We next turn to a comparison of the inferred stellar masses with the
galaxy mass function.  The cumulative distribution of stellar masses
for the short GRB hosts is shown in Figure~\ref{fig:masses_2} with the
range of possible masses bounded by the single age SSP and maximal
values.  We also present a breakdown of the sample into early- and
late-type galaxies, each spanning the same range.  For the late-type
hosts, the intermediate young+old values are also shown.  To compare
these distributions to the distribution of galaxy masses we also plot
the cumulative stellar mass function {\it weighted by mass}, i.e., the
fraction of stellar mass in galaxies above some mass, $f(>M)$, given
by the equation:
\begin{equation}
f(>M)=\frac{\int_M^\infty{{M'\times\Phi(M')\,\,{\rm d}M'}}}
{\int_0^\infty{{M'\times\Phi(M')\,\,{\rm d}M'}}}
\end{equation}
where $\Phi(M)$ is the Schechter mass function:
\begin{equation}
\Phi(M)=\Phi^*\bigg(\frac{M}{M^*}\bigg)^\alpha{\rm
exp}\bigg(-\frac{M}{M^*}\bigg).
\end{equation} 

We use several determinations of $\Phi(M)$ in this comparison,
including the \citet{cnb+01} mass function from the 2MASS/2dF catalogs
for all galaxy types at $z\sim 0$ ($M^*=10^{11.16}$ M$_\odot$,
$\alpha=-1.18$); the nearly identical \citet{phj04} mass function from
SDSS for all galaxy types at $z\sim 0$ ($M^*=10^{11.19}$ M$_\odot$,
$\alpha=-1.16$); the \citet{bmk+03} mass function for late-type
galaxies from 2MASS/SDSS converted to a Salpeter IMF for comparison
with our inferred values ($M^*=10^{10.97}$ M$_\odot$, $\alpha=-1.27$);
and the \citet{isl+10} mass functions from the COSMOS survey for
quiescent galaxies at $z\sim 0.3$ ($M^*=10^{11.13}$ M$_\odot$,
$\alpha=-0.91$) and intermediate-activity galaxies at $z\sim 0.5$
($M^*=10^{10.93}$ M$_\odot$, $\alpha=-1.02$), matched to the redshifts
of the early- and late-type short GRB host galaxies in our sample.
The resulting distributions of $f(>M)$ for the various mass functions
are shown in Figure~\ref{fig:masses_2}.  The median of each
distribution is close to the value of $M^*$, but also depends on the
value of $\alpha$ (\citealt{cnb+01}: $\langle M^*\rangle=10^{10.88}$
M$_\odot$; \citealt{phj04}: $\langle M^*\rangle=10^{10.92}$ M$_\odot$;
\citealt{bmk+03}: $\langle M^*\rangle=10^{10.61}$ M$_\odot$;
\citealt{isl+10}: $\langle M^*\rangle=10^{10.76}$ M$_\odot$ for
intermediate-activity and $\langle M^*\rangle=10^{11.02}$ M$_\odot$
for quiescent).

In \S\ref{sec:res} we demonstrated that the median single age SSP mass
for the short GRB host sample is $1.3\times 10^{10}$ M$_\odot$,
or\footnotemark\footnotetext{In the conversion to units of $\langle
M^*\rangle$ we use the mass function appropriate to each galaxy type,
or the combined sample.} $\approx 0.2\langle M^*\rangle$, well below
the expected value if the short GRB rate tracked stellar mass.  The
result is even more discrepant if we restrict the sample to only the
late-type hosts, with a median of about $0.06\langle M^*\rangle$.  On
the other hand, for the early-type hosts we find a median of about
$1.6\langle M^*\rangle$, suggesting that they are drawn from the
early-type galaxy mass function.

Using the maximal masses for the full sample we find a median stellar
mass of about $1.2\langle M^*\rangle$, while for the subset of
late-type hosts the median is about $0.5\langle M^*\rangle$; the
maximal masses of the early-type hosts lead to a large median mass of
about $5\langle M^*\rangle$, suggesting that they are likely
over-estimates of the true stellar mass.  Finally, using the young+old
stellar masses for the late-type hosts the median mass is about
$0.3\langle M^*\rangle$.  Thus, we find that the distribution of
stellar masses for the late-type hosts is shifted to a somewhat lower
value than the expected median of the galaxy mass function, while the
early-type hosts have a similar median to the galaxy mass function.

Beyond the comparison of median values, we further assess the
agreement (or lack thereof) between the short GRB hosts and galaxy
mass functions using the Kolmogorov-Smirnov (K-S) test.  For the full
sample there is negligible probability that the distribution of single
age SSP masses is drawn from the galaxy mass function, with $P\approx
8\times 10^{-5}$.  On the other hand, for the maximal mass
distribution the probability is $P\approx 0.6$ indicating that for
these masses the short GRB sample is fully consistent with the galaxy
mass function.  Using the intermediate case of SSP masses for the
early-type hosts and the young+old masses for the late-type hosts, we
find a probability of about 0.3, indicating that this combination is
also fully consistent with the galaxy mass function.

Separating the early-type hosts we find that their SSP masses are
fully consistent with the \citet{isl+10} mass function of quiescent
galaxies ($P\approx 0.8$); their large maximal masses, on the other
hand, are inconsistent with the mass function, with $P\approx 0.007$.
Finally, for the late-type hosts we find a clear inconsistency of the
single age SSP masses with the \citet{isl+10} mass function of
intermediate-activity galaxies, with $P\approx 4\times 10^{-7}$.
However, the maximal mass distribution is fully consistent with the
mass function ($P\approx 0.3$), while the young+old mass distribution
is marginally consistent ($P\approx 0.1$).

To summarize, the distribution of short GRB host masses is compatible
with the overall mass distribution of galaxies only if their stellar
masses are given by the SSP masses for the early-type hosts and the
maximal or young+old masses for the late-type hosts.  Since the
opposite scenario (maximal masses for the early-type hosts and SSP
masses for the late-type hosts) is unlikely, we conclude that the
existing sample of short GRB hosts is consistent with the galaxy mass
function.  Equivalently, this means that short GRBs may indeed track
stellar mass alone.  However, we caution that the host demographics
seem to be at odds with the expected equal fractions of total stellar
mass in early- and late-type galaxies, unless nearly all of the
unidentified hosts are early-type galaxies.  This, along with the
somewhat lower than expected masses of the late-type hosts, leaves
open the possibility that at least a subset of short GRB progenitors
track star formation activity rather than stellar mass.

\subsection{Comparison to Long GRBs}
\label{sec:long}

Despite the possibility that some short GRB progenitors may track star
formation activity, we find that the inferred stellar masses and
population ages of short GRB hosts are generally distinct from those
of long GRB hosts in both the single age SSP and maximal models.  Most
importantly, this is true for the subset of late-type hosts.  In the
framework of the single age SSP model we find a K-S probability of
only $0.006$ that the long and short GRB hosts are drawn from the same
mass distribution.  The probability is higher for the subset of
late-type short GRB hosts, $P\approx 0.1$
(Figure~\ref{fig:masses_1a}).  However, since the SSP values represent
the mass of only the young stellar populations, they are mostly
reflective of the star formation activity rather than the total
stellar mass.  If we use instead the maximal masses, the K-S
probability that the long GRB hosts and late-type hosts of short GRBs
are drawn from the same sample is negligible, $P\approx 4\times
10^{-5}$ (Figure~\ref{fig:masses_1b}), demonstrating that they are
distinct galaxy populations.  A similar conclusion is apparent from a
comparison of the single age SSP population ages.  The K-S probability
that the long GRB hosts and late-type hosts of short GRBs are drawn
from the same distribution is only $P\approx 0.006$.

Thus, the long GRB hosts have significantly lower stellar masses than
the subset of late-type short GRB hosts, and their young stellar
population are significantly younger.  Indeed, a comparison of the
long GRB host maximal masses to the \citet{isl+10} mass function of
high-activity galaxies at $z\sim 0.7$ (appropriate for the long GRB
sample considered here) indicates a K-S probability of only $0.002$
that the long GRB hosts are drawn from the galaxy mass function.  This
is consistent with our understanding that their massive star
progenitors select galaxies by star formation (and perhaps additional
factors such as metallicity).

The apparent distinction between long GRB hosts and the late-type
hosts of short GRBs in terms of their stellar masses and young
population ages strengthens our similar previous conclusion based on
their star formation rates, specific star formation rates,
luminosities, and metallicities \citep{ber09}, as well as physical
sizes \citep{fbf10}.  In essentially every property the late-type
short GRB hosts point to a population of more quiescent, massive, and
evolved galaxies than the hosts of long GRBs.  We conclude that this
rules out the idea that short GRB progenitors in late-type hosts are
massive stars identical to long GRB progenitors \citep{vzo+09}, {\it
even if the short GRBs in late-type galaxies indeed track star
formation rather than stellar mass.}

\subsection{The Delay Time Distribution}
\label{sec:delay}

A determination of the delay function from the derived stellar
population ages is complicated by two primary factors.  First, we have
to assume that the short GRB progenitors in each host were formed
within the inferred stellar population.  This assumption is justified
statistically both for an association of the progenitors with stellar
mass and with star formation activity, as long as we can appropriately
normalize the rates of short GRBs.  Second, while we can determine
single age SSP ages from the broad-band photometry, these data are not
sufficient to provide an age breakdown (by mass) for multiple stellar
components.  Indeed, for our hybrid young+old model we had to fix the
age of the old population (to the age of the universe, in this case).
Still, we find that in the young+old model, the bulk of the mass
($\approx 55-99\%$; Table~\ref{tab:short}) is contained in the old
stellar population, and so the progenitors would have ``old'' ages
($\tau\gtrsim\tau_{\rm SSP}$) if they tracked stellar mass.

As a result of these limitations we can only explore the implications
of two main scenarios, namely that short GRBs track mass and/or star
formation activity.  In the context of the former scenario we have
shown in \S\ref{sec:mass} that the short GRBs in early-type hosts
trace stellar mass.  Therefore, we can use their SSP ages to provide a
rough estimate of the progenitor ages, which we find to be
$\tau\approx 0.8-4.4$ Gyr, with a median of about 3 Gyr.  On the other
hand, for the late-type hosts (for which we can extract no credible
information on the mass-weighted stellar population age), we can infer
a typical delay relative to the most recent star formation episode
under the assumption that these progenitors track star formation
activity.  We find SSP ages of $\tau\approx 0.03-0.5$ Gyr, with a
median of about 0.25 Gyr, or young+old ages of about $0.01-0.4$ Gyr
with a median of about 0.1 Gyr.

Thus, if short GRBs follow both stellar mass (in early-type galaxies)
and star formation activity (in late-type galaxies), the typical delay
times are about 3 and 0.2 Gyr, respectively.

\section{Conclusions}
\label{sec:conc}

Using multi-band photometry for 19 short GRB host galaxies we
extracted stellar masses and ages under the assumption of single age
populations, combined young+old populations, and maximal mass
populations.  The resulting values allow us to investigate whether
short GRBs track the cosmic stellar mass distribution, to estimate
their typical delay times if mass and/or star formation play a role,
and to investigate whether short GRBs in late-type galaxies share
progenitors with long GRBs.  The basic result is that the early-type
host SSP masses track the mass distribution of early-type galaxies,
while for the late-type hosts this is only the case if they have
maximal or possibly young+old masses.  However, the host demographics
do not appear to follow the expected mass-weighted one-to-one ratio,
unless nearly all of the current hosts of unknown type are early-type
galaxies.  We note that additional support for a short GRB rate
tracking stellar mass is provided by high resolution {\it Hubble Space
Telescope} ({\it HST}) imaging \citep{fbf10}, which indicates that the
locations of these bursts better track stellar mass (rest-frame
optical light) than star formation (rest-frame UV light).

We further find that the late-type hosts of short GRBs and the hosts
of long GRBs are distinct in their stellar masses and young population
ages, with the former having higher masses by about an order of
magnitude and older ages by about a factor of 4.  This result is in
good agreement with the previous conclusion that short GRB hosts have
lower star formation rates and higher metallicities compared to long
GRB hosts \citep{ber09}.

These observational results lead to several crucial conclusions
regarding the nature of short GRB progenitors:

\begin{itemize}

\item Short GRB progenitors are generally consistent with arising in
old stellar populations that track the cosmic stellar mass
distribution.

\item Despite this overall consistency, the dearth of early-type
hosts, and the somewhat lower than expected masses of the late-type
hosts indicate that short GRB progenitors in late-type galaxies may
partially track star formation activity.

\item If short GRB progenitors track only stellar mass, then the
typical stellar population ages of the early-type hosts indicate a
typical progenitor age (or delay time) of $\sim 3$ Gyr (with a range
of $\sim 1-5$ Gyr).  If the progenitors in late-type hosts track star
formation activity, the resulting typical delay is about 0.2 Gyr.

\item Even if short GRBs in late-type galaxies track star formation
activity, their delay times are at least a few times longer than for
long GRBs, indicating that they do not share the same progenitors.
This is also supported by the substantially different mass
distributions of long and short GRB hosts.

\end{itemize}

The conclusion that short GRBs are consistent with tracking stellar
mass provides additional support for the compact object merger model,
although it does not rule out other progenitors with substantial time
delays (e.g., binary white dwarf mergers, or white dwarf and neutron
star accretion-induced collapse).  In the context of these various
models, it is possible that a short delay channel exists, but a
substantial population with delays of only a few Myr or even tens of
Myr (e.g., \citealt{bpb+06}) is not supported by the data.

The clear distinction between short and long GRB hosts, and the
inferred delay times, call into question the results of previous
analyses claiming a substantial overlap between long and short GRB
progenitors.  For example, it demonstrates that blunt statistical
tools such as the ${\rm log}N-{\rm log}S$ distribution \citep{vzo+09}
do not have sufficient time resolution to distinguish between truly
young populations such as long GRB massive star progenitors, and
intermediate-age progenitors as inferred here if short GRBs in
late-type galaxies track star formation activity.  Similarly, these
results call into question a recent proposal that some short GRBs are
due to widely off-axis ``cocoon'' emission from long GRBs
\citep{lmb10}.

Our results also demonstrate that caution should be taken with the
proposed re-classification of short and long GRBs into Type I and II
events, marking old and young progenitors, respectively
\citep{zzl+07,zzv+09}.  Such a new bimodal classification may lead to
the erroneous conclusion that short GRBs in late-type galaxies (even
if they track the on-going star formation activity) share the same
progenitors as long GRBs (e.g., \citealt{vzo+09}) since both would be
classified as Type II.  At the very least, such a new classification
scheme may require a further breakdown of the Type II events into
those that result from massive stars versus those that simply track
star formation activity with a modest delay, e.g., Type IIa and IIb.
Clearly, this is beyond the scope of the current short GRB sample.  We
note, by analogy, that for Type Ia supernovae a connection with both
stellar mass and star formation activity has been demonstrated
\citep{mdp06,slp+06,atj+08,bta+10}, but the prompt channel is not
related to the massive star progenitors of Type II and Ib/c
supernovae.

We conclude by noting that two primary refinements of our analysis
technique can be made with future observations.  First, high
signal-to-noise ratio spectra can be used to extract more detailed
star formation histories than the available broad-band photometry.
Such reconstructed histories will allow us to more robustly estimate
the delay time distribution in the context of a mass-weighted rate
(c.f., \citealt{atj+08} for Type Ia supernovae).  Second, high angular
resolution multi-band imaging with {\it HST} will allow us to
determine spatially-resolved ages at the locations of short GRBs and
hence to further refine the connection with young or old populations.
Both of these techniques are within our grasp in the near future.

\acknowledgements We acknowledge helpful discussions with J.~Strader,
E.~Westra, R.~Foley, R.~Chornock, E.~Nakar, E.~Levesque, and
A.~Soderberg.  This paper includes data gathered with the 6.5 meter
Magellan Telescopes located at Las Campanas Observatory, Chile.  It is
also based in part on observations obtained at the Gemini Observatory,
which is operated by the Association of Universities for Research in
Astronomy, Inc., under a cooperative agreement with the NSF on behalf
of the Gemini partnership: the National Science Foundation (United
States), the Particle Physics and Astronomy Research Council (United
Kingdom), the National Research Council (Canada), CONICYT (Chile), the
Australian Research Council (Australia), CNPq (Brazil) and CONICET
(Argentina).  This work was partially supported by Swift AO5 grant
\#5080010 and AO6 grant \#6090612.  Additional support was provided by
the Harvard College Research Program.


\clearpage
\LongTables
\begin{deluxetable}{lllccccll}
\tabletypesize{\footnotesize}
\tablecolumns{9} 
\tabcolsep0.1in\footnotesize
\tablewidth{0pt} 
\tablecaption{Short GRB Host Galaxy Photometry
\label{tab:data}}
\tablehead{
\colhead{GRB}      &
\colhead{RA}       &
\colhead{Dec}      &
\colhead{$z$}      &
\colhead{OA?}      &
\colhead{Type$\,^a$} &
\colhead{Filter}     &
\colhead{$m_{AB}\,^b$} &
\colhead{Refs.}    \\
\colhead{}         &        
\colhead{(J2000)}  &
\colhead{(J2000)}  &       
\colhead{}         & 
\colhead{}         & 
\colhead{}         & 
\colhead{}         & 
\colhead{(mag)}    & 
\colhead{}                                          
}
\startdata
050509b & \ra{12}{36}{13.58} & \dec{+28}{59}{01.3}   & 0.225      & N & E & u & $20.22\pm 0.13$ & SDSS  \\
        &                    &                       &            &   &   & g & $18.45\pm 0.02$ & SDSS  \\
        &                    &                       &            &   &   & r & $17.07\pm 0.01$ & SDSS  \\
        &                    &                       &            &   &   & i & $16.56\pm 0.01$ & SDSS                \\  
        &                    &                       &            &   &   & z & $16.22\pm 0.01$ & SDSS                 \\
        &                    &                       &            &   &   & J & $16.14\pm 0.14$ & 2MASS                  \\
        &                    &                       &            &   &   & H & $15.83\pm 0.18$ & 2MASS                 \\
        &                    &                       &            &   &   & K & $15.97\pm 0.16$ & 2MASS                 \\
050709  & \ra{23}{01}{26.96} & \dec{$-$38}{58}{39.5} & 0.161      & Y & L & B & $22.00\pm 0.10$ & \citet{hwf+05}                   \\
        &                    &                       &            &   &   & V & $21.31\pm 0.07$ & \citet{cmi+06}                  \\
        &                    &                       &            &   &   & R & $21.23\pm 0.07$ & \citet{cmi+06}                  \\
        &                    &                       &            &   &   & I & $20.99\pm 0.08$ & \citet{cmi+06}                  \\  
        &                    &                       &            &   &   & J & $20.75\pm 0.08$ & This Work                \\
        &                    &                       &            &   &   & K & $21.04\pm 0.16$ & This Work                  \\
050724  & \ra{16}{24}{44.36} & \dec{$-$27}{32}{27.5} & 0.257      & Y & E & U & $>23.09$        & \citet{gcg+06} \\         
        &                    &                       &            &   &   & B & $19.69\pm 0.12$ & \citet{gcg+06}                 \\
        &                    &                       &            &   &   & V & $18.66\pm 0.05$ & \citet{gcg+06}                  \\
        &                    &                       &            &   &   & R & $18.19\pm 0.03$ & \citet{gcg+06}                   \\
        &                    &                       &            &   &   & I & $17.82\pm 0.20$ & \citet{bpc+05}               \\  
        &                    &                       &            &   &   & J & $17.28\pm 0.04$ & \citet{gcg+06}                  \\
        &                    &                       &            &   &   & H & $16.89\pm 0.05$ & \citet{gcg+06}                  \\ 
        &                    &                       &            &   &   & K & $16.59\pm 0.05$ & \citet{gcg+06}                 \\
051210  & \ra{22}{00}{41.26} & \dec{$-$57}{36}{46.5} & \nod       & N & ? & g & $24.22\pm 0.34$ & This Work                 \\
        &                    &                       &            &   &   & r & $23.99\pm 0.15$ & This Work                 \\
        &                    &                       &            &   &   & i & $24.86\pm 0.22$ & This Work                 \\  
        &                    &                       &            &   &   & z & $24.03\pm 0.21$ & This Work                 \\      
        &                    &                       &            &   &   & K & $>20.9$         & This Work                 \\     
051221a & \ra{21}{54}{48.62} & \dec{+16}{53}{27.2}   & 0.546      & Y & L & g & $23.48\pm 0.07$ & his Work                 \\
        &                    &                       &            &   &   & r & $21.99\pm 0.09$ & \citet{sbk+06}                  \\
        &                    &                       &            &   &   & i & $21.99\pm 0.17$ & \citet{sbk+06}      \\  
        &                    &                       &            &   &   & z & $21.97\pm 0.40$ & \citet{sbk+06}                 \\
        &                    &                       &            &   &   & J & $21.95\pm 0.20$ & This Work                  \\
        &                    &                       &            &   &   & K & $22.27\pm 0.15$ & This Work                   \\
060801  & \ra{14}{12}{01.35} & \dec{+16}{58}{53.7}   & 1.130      & N & L & g & $23.37\pm 0.09$ & This Work                 \\
        &                    &                       &            &   &   & r & $23.15\pm 0.11$ & This Work                   \\
        &                    &                       &            &   &   & i & $23.01\pm 0.19$ & This Work                  \\  
        &                    &                       &            &   &   & z & $22.85\pm 0.10$ & This Work                   \\   
        &                    &                       &            &   &   & J & $>21.5$         & This Work                   \\  
        &                    &                       &            &   &   & K & $>19.9$         & This Work                   \\   
061006  & \ra{07}{24}{07.66} & \dec{$-$79}{11}{55.1} & 0.438      & Y & L & B & $24.38\pm 0.12$ & \citet{dmc+09}                  \\
        &                    &                       &            &   &   & V & $23.51\pm 0.07$ & \citet{dmc+09}                  \\
        &                    &                       &            &   &   & r & $23.28\pm 0.09$ & This Work                \\
        &                    &                       &            &   &   & R & $23.29\pm 0.12$ & \citet{dmc+09}                 \\
        &                    &                       &            &   &   & I & $22.82\pm 0.12$ & \citet{dmc+09}                  \\ 
        &                    &                       &            &   &   & z & $22.81\pm 0.25$ & This Work                  \\  
        &                    &                       &            &   &   & J & $22.62\pm 0.20$ & \citet{dmc+09}                  \\
        &                    &                       &            &   &   & K & $22.48\pm 0.25$ & This Work                 \\
061210  & \ra{09}{38}{05.27} & \dec{+15}{37}{17.3}   & 0.410      & N & L & g & $23.15\pm 0.10$ & This Work                  \\
        &                    &                       &            &   &   & r & $21.30\pm 0.05$ & This Work                 \\
        &                    &                       &            &   &   & i & $21.60\pm 0.10$ & This Work                  \\  
        &                    &                       &            &   &   & z & $21.27\pm 0.14$ & This Work                  \\
        &                    &                       &            &   &   & J & $21.29\pm 0.15$ & This Work                  \\
        &                    &                       &            &   &   & K & $20.32\pm 0.10$ & This Work                  \\
061217  & \ra{10}{41}{39.32} & \dec{$-$21}{07}{22.1} & 0.827      & N & L & g & $23.30\pm 0.10$ & This Work                 \\
        &                    &                       &            &   &   & r & $22.92\pm 0.10$ & This Work                  \\
        &                    &                       &            &   &   & i & $22.42\pm 0.06$ & This Work                  \\  
        &                    &                       &            &   &   & z & $22.39\pm 0.08$ & This Work                  \\
        &                    &                       &            &   &   & J & $23.17\pm 0.35$ & This Work                  \\
        &                    &                       &            &   &   & K & $23.06\pm 0.23$ & This Work                \\
070429b & \ra{21}{52}{03.84} & \dec{$-$38}{49}{42.4} & 0.902      & N & L & g & $24.30\pm 0.20$ & This Work                 \\
        &                    &                       &            &   &   & r & $23.21\pm 0.04$ & This Work                  \\
        &                    &                       &            &   &   & i & $21.84\pm 0.09$ & This Work                   \\  
        &                    &                       &            &   &   & z & $21.72\pm 0.12$ & This Work                 \\             
070714b & \ra{03}{51}{22.30} & \dec{+28}{17}{50.8}   & 0.923      & Y & L & g & $25.23\pm 0.34$ & \citet{gfl+09}                 \\
        &                    &                       &            &   &   & r & $24.50\pm 0.21$ & \citet{gfl+09}                 \\
        &                    &                       &            &   &   & i & $23.67\pm 0.12$ & \citet{gfl+09}                 \\  
        &                    &                       &            &   &   & z & $23.77\pm 0.13$ & \citet{gfl+09}                 \\
        &                    &                       &            &   &   & J & $23.05\pm 0.12$ & \citet{gfl+09}                  \\
        &                    &                       &            &   &   & H & $23.58\pm 0.20$ & \citet{gfl+09}                  \\
        &                    &                       &            &   &   & K & $22.97\pm 0.13$ & \citet{gfl+09}                  \\
070724  & \ra{01}{51}{13.96} & \dec{$-$18}{35}{40.1} & 0.457      & Y & L & g & $21.51\pm 0.06$ & This Work                  \\
        &                    &                       &            &   &   & r & $20.74\pm 0.03$ & This Work                  \\
        &                    &                       &            &   &   & i & $20.43\pm 0.03$ & This Work                 \\  
        &                    &                       &            &   &   & z & $20.26\pm 0.04$ & This Work                 \\
        &                    &                       &            &   &   & J & $20.01\pm 0.02$ & This Work                  \\
        &                    &                       &            &   &   & H & $19.78\pm 0.02$ & This Work                  \\
        &                    &                       &            &   &   & K & $19.71\pm 0.04$ & This Work                   \\
070729  & \ra{03}{45}{16.04} & \dec{$-$39}{19}{19.9} & \nod       & N & ? & g & $24.38\pm 0.38$ & This Work                 \\
        &                    &                       &            &   &   & r & $23.31\pm 0.03$ & This Work                 \\
        &                    &                       &            &   &   & i & $21.80\pm 0.06$ & This Work                  \\  
        &                    &                       &            &   &   & z & $21.82\pm 0.17$ & This Work                  \\
        &                    &                       &            &   &   & J & $20.85\pm 0.10$ & This Work                  \\
        &                    &                       &            &   &   & K & $20.10\pm 0.10$ & This Work                  \\
070809$\,^c$ & \ra{13}{35}{04.41} & \dec{$-$22}{08}{28.9} & 0.473 & Y & E & g & $21.80\pm 0.05$ & This Work                  \\
        &                    &                       &            &   &   & r & $19.89\pm 0.02$ & This Work                  \\
        &                    &                       &            &   &   & i & $19.27\pm 0.05$ & This Work                  \\  
        &                    &                       &            &   &   & K & $17.96\pm 0.04$ & This Work                  \\
071227  & \ra{03}{52}{31.26} & \dec{$-$55}{59}{03.5} & 0.381      & Y & L & g & $22.82\pm 0.13$ & This Work                 \\
        &                    &                       &            &   &   & r & $20.60\pm 0.05$ & This Work                 \\
        &                    &                       &            &   &   & i & $20.47\pm 0.04$ & This Work                  \\  
        &                    &                       &            &   &   & z & $19.77\pm 0.03$ & This Work                  \\
        &                    &                       &            &   &   & J & $19.16\pm 0.06$ & This Work                  \\
        &                    &                       &            &   &   & K & $18.15\pm 0.06$ & This Work                  \\
080123  & \ra{22}{35}{46.10} & \dec{$-$64}{54}{03.2} & 0.495      & N & L & g & $22.06\pm 0.06$ & This Work                  \\
        &                    &                       &            &   &   & r & $20.89\pm 0.05$ & This Work                  \\
        &                    &                       &            &   &   & i & $20.49\pm 0.07$ & This Work                  \\  
        &                    &                       &            &   &   & z & $20.12\pm 0.20$ & This Work                  \\
        &                    &                       &            &   &   & J & $20.30\pm 0.05$ & This Work                  \\
        &                    &                       &            &   &   & K & $19.58\pm 0.06$ & This Work                  \\
090510  & \ra{22}{14}{12.56} & \dec{$-$26}{34}{59.0} & 0.903      & Y & L & g & $23.78\pm 0.08$ & This Work                  \\
        &                    &                       &            &   &   & i & $22.41\pm 0.14$ & This Work                  \\  
        &                    &                       &            &   &   & z & $22.66\pm 0.17$ & This Work                  \\
        &                    &                       &            &   &   & J & $21.79\pm 0.15$ & This Work                   \\
        &                    &                       &            &   &   & K & $21.98\pm 0.35$ & This Work                  \\
090515$\,^c$ & \ra{10}{56}{36.11} & \dec{+14}{26}{30.3}   & 0.403 & Y & E & g & $21.89\pm 0.02$ & This Work                   \\
        &                    &                       &            &   &   & r & $20.21\pm 0.05$ & This Work                   \\
        &                    &                       &            &   &   & i & $19.45\pm 0.05$ & This Work                  \\  
        &                    &                       &            &   &   & J & $18.67\pm 0.05$ & This Work                  \\
        &                    &                       &            &   &   & K & $18.23\pm 0.05$ & This Work                   \\
100117  & \ra{00}{45}{04.7}  & \dec{$-$01}{35}{42.0} & 0.920      & Y & E & g & $26.01\pm 0.30$ & Fong et al.~in prep.       \\
        &                    &                       &            &   &   & r & $23.97\pm 0.04$ & Fong et al.~in prep.        \\
        &                    &                       &            &   &   & J & $21.72\pm 0.26$ & Fong et al.~in prep.         \\
        &                    &                       &            &   &   & H & $21.63\pm 0.21$ & Fong et al.~in prep.        \\
        &                    &                       &            &   &   & K & $21.24\pm 0.25$ & Fong et al.~in prep.        
\enddata

\tablecomments{Properties and photometric measurements for short GRB
host galaxies in this paper.\\
$^a$ Type indicates whether a galaxy is early-type (E) or late-type
(L).\\
$^b$ Corrected for Galactic extinction \citep{sfd98}.\\
$^c$ These bursts do not have coincident host galaxies.  Magnitudes
are provided for the galaxy with the lowest probability of chance
coincidence \citep{ber10}.}
\end{deluxetable}

\clearpage
\begin{deluxetable}{lcccccccccccc}
\tabletypesize{\footnotesize}
\tablecolumns{13} 
\tabcolsep0.08in\footnotesize
\tablewidth{0pc} 
\tablecaption{Stellar Ages and Masses of Short GRB Host Galaxies
\label{tab:short}}
\tablehead{
\colhead{}               &
\colhead{}               &
\colhead{}               &
\multicolumn{3}{c}{SSP$\,^a$}  &
\colhead{}               &
\multicolumn{5}{c}{Young+Old$\,^b$}  &
\colhead{}               \\
\cline{4-6}              
\cline{8-12}             \\
\colhead{GRB}            &        
\colhead{$z$}            &
\colhead{Type}           &
\colhead{$\tau$}         &
\colhead{${\rm log}(M)$} &
\colhead{$\chi^2_r$}     &
\colhead{}               &
\colhead{$\tau_{\rm Y}$}         &
\colhead{${\rm log}(M_{\rm Y})$} &
\colhead{$\tau_{\rm O}\,^c$}     &
\colhead{${\rm log}(M_{\rm O})$} &
\colhead{$\chi^2_r$}             &
\colhead{${\rm log}(M_{\rm max})\,^d$} \\
\colhead{}               &        
\colhead{}               &
\colhead{}               &
\colhead{(Gyr)}          &
\colhead{(M$_\odot$)}    &
\colhead{}               &
\colhead{}               &
\colhead{(Gyr)}          &
\colhead{(M$_\odot$)}    &
\colhead{(Gyr)}          &
\colhead{(M$_\odot$)}    &
\colhead{}               &
\colhead{(M$_\odot$)}    
}
\startdata
050509b      & 0.225            & E & 3.18 & 11.6 & 0.7 & & \nod  & \nod & \nod & \nod    & \nod & 11.9 \\
050709       & 0.161            & L & 0.26 & 8.8  & 0.6 & & 0.21  & 8.7  & 11.6 & 9.2     & 0.6  & 9.7  \\
050724       & 0.257            & E & 0.94 & 10.8 & 0.3 & & \nod  & \nod & \nod & \nod    & \nod & 11.7 \\
051210       & $1.3\pm 0.3\,^e$ & ? & 0.03 & 8.8  & 2.5 & & \nod  & \nod & \nod & \nod    & \nod & \nod \\
051221       & 0.547            & L & 0.17 & 9.4  & 2.7 & & 0.14  & 9.3  & 8.3  & $<9.4$  & 2.4  & 10.1 \\
060801       & 1.130            & L & 0.03 & 9.1  & 1.2 & & 0.03  & 9.1  & 5.4  & $<10.8$ & 1.9  & 10.9 \\
061006       & 0.438            & L & 0.24 & 9.0  & 1.3 & & 0.09  & 8.5  & 9.1  & 9.7     & 0.8  & 9.8  \\
061210       & 0.410            & L & 0.38 & 9.6  & 4.2 & & 0.38  & 9.5  & 9.1  & $<9.9$  & 5.2  & 10.5 \\
061217       & 0.827            & L & 0.03 & 9.1  & 1.7 & & 0.03  & 9.0  & 6.6  & $<9.3$  & 2.1  & 10.1 \\
070429b      & 0.902            & L & 0.46 & 10.4 & 1.8 & & 0.35  & 10.2 & 6.3  & $<11.0$ & 2.6  & 11.3 \\
070714b      & 0.923            & L & 0.22 & 9.4  & 2.9 & & 0.10  & 9.1  & 6.2  & $<10.2$ & 3.3  & 10.2 \\
070724       & 0.457            & L & 0.30 & 10.1 & 2.1 & & 0.01  & 8.3  & 8.9  & 10.9    & 0.3  & 11.0 \\
070729       & $0.8\pm 0.1\,^e$ & ? & 0.98 & 10.6 & 2.7 & & \nod  & \nod & \nod & \nod    & \nod & \nod \\ 
070809       & 0.473            & E & 3.07 & 11.4 & 0.9 & & \nod  & \nod & \nod & \nod    & \nod & 11.7 \\
071227$\,^f$ & 0.381            & L & 0.49 & 10.4 & 2.2 & & 0.36  & 10.3 & 9.3  & $<11.0$ & 2.7  & 11.4 \\    
080123       & 0.495            & L & 0.31 & 10.1 & 1.7 & & 0.03  & 9.7  & 8.7  & $<10.7$ & 1.8  & 11.1 \\
090510       & 0.903            & L & 0.14 & 9.7  & 3.3 & & 0.02  & 9.3  & 6.3  & $<10.4$ & 3.7  & 10.6 \\
090515       & 0.403            & E & 4.35 & 11.2 & 0.1 & & \nod  & \nod & \nod & \nod    & \nod & 11.5 \\
100117       & 0.920            & E & 0.79 & 10.3 & 1.8 & & \nod  & \nod & \nod & \nod    & \nod & 10.9
\enddata
\tablecomments{Stellar population parameters derived from the
broad-band SEDs in Table~\ref{tab:data} using the \citet{mar05}
stellar population synthesis models (\S\ref{sec:model}).\\
$^a$ Fit with a single stellar population age. \\
$^b$ Fit with a two-component young+old SSP model, with the old
population fixed at an age of $\tau_{\rm O}$ equivalent to the age of
the universe at each host redshift, and a freely-varying young
population with an age of $\tau_{\rm Y}$. \\
$^c$ $\tau_{\rm O}$ is fixed at the age of the universe at the
appropriate redshift.\\
$^d$ The maximal mass is derived using the observed $K$-band fluxes
assuming a stellar population age equal to the age of the universe at
the host redshift.\\
$^e$ A spectroscopic redshift is not known; these values are derived
photometrically from the \citet{mar05} models.\\
$^f$ This model includes host extinction with $A_{\rm V,host}=1.5$
mag; a model with no host extinction does not provide a satisfactory
fit.} 
\end{deluxetable}

\clearpage
\begin{deluxetable}{llcccc}
\tabletypesize{\footnotesize}
\tablecolumns{6} 
\tabcolsep0.12in\footnotesize
\tablewidth{0pt} 
\tablecaption{Stellar Ages and Masses of Long GRB Host Galaxies
\label{tab:long}}
\tablehead{
\colhead{}                     &
\colhead{}                     &
\multicolumn{3}{c}{SSP$\,^a$}  &
\colhead{}                     \\
\cline{3-5}                    \\
\colhead{GRB}                  &
\colhead{$z$}                  &
\colhead{Age}                  &
\colhead{Mass}                 &
\colhead{$\chi^2_r$}           &
\colhead{${\rm log}(M_{\rm max})\,^b$} \\
\colhead{}                     &        
\colhead{}                     &              
\colhead{(Gyr)}                &            
\colhead{(M$_\odot$)}          &
\colhead{}                     &              
\colhead{(M$_\odot$)}          
}            
\startdata
970228  & 0.695 & 0.14       & 8.6  & 1.1 & 8.7  \\
970508  & 0.835 & 0.07       & 8.5  & 0.4 & 8.8  \\
980613  & 1.097 & 0.05       & 9.1  & 3.6 & 9.4  \\
        &       & 0.02$\,^c$ & 9.0  & 1.0 &      \\
980703  & 0.966 & 0.05       & 9.3  & 1.0 & 10.8 \\
990123  & 1.600 & 0.05       & 9.4  & 2.0 & 9.7  \\
990712  & 0.433 & 0.11       & 9.1  & 1.1 & 9.3  \\
991208  & 0.706 & 0.10       & 8.8  & 0.6 & 9.0  \\
000210  & 0.846 & 0.12       & 9.3  & 0.7 & 9.5  \\
000418  & 1.118 & 0.05       & 9.3  & 2.0 & 9.6  \\
000911  & 1.058 & 0.06       & 8.9  & 0.2 & 9.6  \\
010222  & 1.480 & 0.06       & 8.6  & 0.8 & 8.9  \\
010921  & 0.451 & 0.22       & 9.6  & 0.4 & 9.6  \\
011121  & 0.362 & 0.24       & 9.5  & 2.0 & 10.3 \\
020813  & 1.255 & 0.09       & 9.4  & 1.3 & 9.6  \\
020819b & 0.410 & 0.35       & 10.3 & 1.0 & 10.5 \\
020903  & 0.251 & 0.01       & 7.8  & 0.9 & 9.1  \\
030328  & 1.520 & 0.06       & 9.3  & 0.7 & 9.6  \\
030329  & 0.168 & 0.01       & 6.8  & 1.1 & 8.0  \\
030528  & 0.782 & 0.01       & 9.0  & 0.4 & 10.0 \\
        &       & 0.04       & 8.6  & 0.7 &      \\
050826  & 0.296 & 0.50       & 9.9  & 2.0 & 11.0 \\
060218  & 0.034 & 0.01       & 6.7  & 0.1 & 9.1  \\
        &       & 0.06       & 7.5  & 0.3 &      \\
060505  & 0.089 & 0.23       & 9.4  & 2.4 & 9.7  \\
        &       & 0.01$\,^d$ & 8.3  & 0.9 &      \\
060614  & 0.125 & 0.57       & 8.1  & 0.3 & 9.2 
\enddata
\tablecomments{Stellar population parameters derived from the
broad-band SEDs using the \citet{mar05} stellar population synthesis
models (\S\ref{sec:model}).\\
$^a$ Fit with a single stellar population age. \\
$^b$ The maximal mass is derived using the observed $K$-band flux in
conjunction with the maximal age allowed by the burst redshift.\\
$^c$ This model includes host extinction with $A_{\rm V,host}=0.5$
mag. \\
$^d$ This model includes host extinction with $A_{\rm V,host}=0.3$
mag. \\
}
\end{deluxetable}

\clearpage

\clearpage

\clearpage

\clearpage

\clearpage

\clearpage

\clearpage

\clearpage

\end{document}